\documentclass{article}

\usepackage{arxiv}
\usepackage{authblk}

\usepackage[utf8]{inputenc} 
\usepackage{hyperref}       
\usepackage{url}            
\usepackage{amsfonts}       
\usepackage{nicefrac}       
\usepackage{microtype}      
\usepackage{lipsum}
\usepackage{graphicx}
\graphicspath{ {./images/} }

\usepackage{algorithm}
\usepackage{algpseudocode}
\usepackage{graphicx}
\usepackage{subcaption}
\usepackage{csquotes} 
\usepackage{wrapfig}
\usepackage{tikz}
\usepackage{multicol}
\usepackage{amsmath,amsthm,amssymb}
\usepackage{float}
\usepackage{soul}
\usepackage{enumitem}
\usepackage{physics}

\usepackage{array}
\usepackage{xr}
\usepackage{titlesec} 

\title{Distance Backbones Optimize Spreading Dynamics and Centrality Ranks in the Sparsification of Complex Networks}

\author[1, 2, 3, $\dagger$]{Bernardo Pereira}
\author[3, 4, $\dagger$, *]{Felipe Xavier Costa}
\author[4, 3, *]{Luís M. Rocha}

\affil[1]{Dipartimento di Scienze Matematiche, Politecnico di Torino, Turin, 10129, Italy}
\affil[2]{CENTAI Institute, Turin, 10138, Italy}
\affil[3]{Universidade Católica Portuguesa, Católica Medical School, Católica Biomedical Research Centre, Portugal}
\affil[4]{School of Systems Science and Industrial Engineering, Binghamton University (State University of New York), Binghamton, NY 13902, USA}
\affil[$\dagger$]{Equal contribution}
\affil[*]{Corresponding authors: FXC (fxcosta@ucp.pt) and LMR (rocha@binghamton.edu)}

\begin{document}

\maketitle

\begin{abstract}

Detailed network models of social, biological and other complex systems are often dense, which increases their computational complexity in simulations and analysis.
To address this challenge, graph sparsification is used to remove edges while preserving desired network properties.
Distance backbones of weighted graphs, which remove edges that break a generalized triangle inequality for any given path-length measure, preserve all shortest paths of weighted graphs. 
They have been shown to typically sparsify graphs more, as well as preserve community structure and spreading dynamics better than alternative state-of-the-art methods. 
Here, We show that they significantly best preserve node centrality ranks, as well as local and global dynamics in spreading phenomena. 
This is done by introducing the \textit{distance backbone synthesis} (DBS) to progressively sparsify weighted graphs according to a general family of nested distance backbones, whereby each edge is associated with the smallest distance backbone in which it appears. 
DBS provides a principled and natural method to sweep all degrees of sparsification possible while preserving connectivity, allowing us to precisely study (directed and undirected) weighted graph sparsification under multi-objective criteria.
It provides an algebraically-principled explanation of edge importance by revealing the precise topological space associated with each edge.
The theory is demonstrated with a battery of social contact networks obtained from real-world social activity in different scenarios.
Our study also shows that the optimal preservation of node centrality and spreading dynamics happens for the distance backbone obeying the generalized triangle inequality for the path-length measure $g(x, y) = (\sqrt[3]{x}+\sqrt[3]{y})^3$, which removes more than half of the edges from the empirical networks studied.

\end{abstract}


\section{Introduction}
\label{sec:introduction}

Network analysis has been increasingly used to uncover complex patterns (central components, communities, etc) in social, biological and many other systems.
For example, in transportation systems, the flux of passengers among different locations can be mapped into a mobility network, which enables the identification of key traffic routes and most active locations towards developing effective traffic policies \cite{colizza2006role,munozmendez2018community}. 
Meanwhile, a social system can be studied as a network of individuals connected by their frequency of contacts to inform disease mitigation strategies via network analysis \cite{starnini2013immunization,toth2015role}.
Generally one has relational information among multiple nodes (locations, individuals, etc) being represented in a graph with many edges (flux of passengers, frequency of contacts, etc) connecting node pairs.
With a growing number of nodes, coming usually from a more detailed system description, so does grow the number of edges which leads to challenges in both computational resources for network analysis and theoretical explainability.
Those challenges call for the development of \emph{network sparsification} methods capable of identifying subgraphs that preserve features of interest from the system's network representation.

The multitude of methods available to tackle network sparsification highlights the complexity of determining which edges are essential and which are redundant for network inference.
A first approach to network sparsification is removing edges with small weights in a network, as they encode for weaker interactions \cite{yan2018weight}.
However, this method often leads to disconnected networks and fails to preserve the original network’s structural and dynamical properties, as weaker edges may play key roles in both characteristics \cite{correia2023contact}.
Another option compares the weights of edges against a null model to determine which edges are statistically significant, a method referred to as the disparity filter \cite{serrano2009extracting}.
By retaining only those edges that don't deviate from the null model, this method allows for a more informed reduction of the network but is biased by the chosen model.
A more unbiased approach is the effective resistance method \cite{spielman2011graph, spielman2011spectral}, which uses spectral graph theory to incorporate information about all parallel paths between pairs nodes, proceeding analogously to the computation of effective resistances in an electrical circuit.
A variant of this methodology, where each effective resistance is weighted \textit{a posteriori} by the original edge weight, has also been proposed \cite{mercier2022effective}.
These methods, however, requires the computation of the pseudo-inverse of the Laplacian matrix, which for large-graphs needs to be done using approximation algorithms. 
Other parametric algorithms using a mix of structural network properties and their statistical significance have been proposed to quantify edges importance for thresholding such as the high-salience skeleton \cite{grady2012robust, shekhtman2014robustness}, omega-based link removal \cite{qiu2024inverse}, and semi-metric distortion sparsification \cite{soriano2025quantifying}.
An exact, algebraically principled, and scalable network sparsification method is the identification of the \emph{metric backbone}, which removes edges that break the triangle inequality and thereby being redundant for shortest-path computation \cite{simas2021distance, costa2023distance}.
Generalizing the triangle inequality, one can even sparsify weighted directed networks until the union of all minimum spanning forests, also called its \emph{ultrametric backbone} \cite{rozum2024ultrametric}.
In this work we explore other backbones \cite{simas2021distance}, which depend on the generalization of the triangle inequality considered \cite{simas2015distance}, while measuring their performance in preserving desired network features.

One of the main goals behind network representation of complex social and biological systems is inferring the importance of its constituent parts, such as nodes and pathways, in information spreading (disease, news, etc).
In this context, it is important that network sparsification methods are capable of preserving spreading dynamics happening in the original network \cite{correia2023contact, soriano2025quantifying}.
Particularly, the effective resistance method has been shown to preserve epidemic spreading under the Susceptible-Infected (SI) \cite{swarup2016identifying} and Susceptible-Infected-Recovered (SIR) \cite{mercier2022effective} dynamics. 
Moreover, when shortest-path redundancy is accounted for via the \emph{metric backbone} and \emph{semi-metric distortion} methods, sparser and more explainable subgraphs are found to preserve SI \cite{correia2023contact} and Susceptible-Infected-Susceptible (SIS) \cite{soriano2025quantifying} dynamics.
That is, the main pathways for information spreading in complex networks are those close to shorter ones, given that the length of a path is found by distance addition. 
Building up on the central role portrayed by removing shortest-paths redundancy to spreading dynamics \cite{correia2023contact, soriano2025quantifying}, this paper introduces a novel method for progressive network sparsification, which we call the \emph{Distance Backbone Synthesis}, capable of identifying the main subgraph for dynamical processes by
ranking edges according to the topological space in which they become a shortest path.
We evaluate our method capabilities to preserve SI spreading dynamics and eigenvector centrality, the latter being an indirect marker of network response to many dynamical processes \cite{jackson2008social, pastor2015epidemic, fletcher2018structure}.
Our results are compared with state-of-the-art network sparsification methods (disparity filter \cite{serrano2009extracting}, weighted effective resistance \cite{mercier2022effective}, and semi-metric distortion sparsification \cite{soriano2025quantifying}) and reveal that the \emph{Distance Backbone Synthesis} explains the topological space on which dynamical processes are occurring via the optimal subgraph preserving the dynamics.
Moreover, we uncover the metric and product backbones \cite{simas2021distance} as non-parametric strict and lenient, respectively, network sparsification bounds.

\section{Background}
\label{sec:background}

A weighted network representation of a complex system, $G$, is made up of a set of nodes, $X$, which are connected according to a measure of \emph{proximity} or \emph{distance} between node pairs depending on what the edge weights quantify.
Proximity edge weights quantify a notion of normalized interaction strength whereby the graph adjacency matrix has elements $p_{ij}\in[0,1]$, with $p_{ij}=0$ denoting that two nodes $x_i, x_j \in X$ are not connected and $p_{ij}=1$ denoting the strongest possible interaction between node pairs.
On the other hand, distance edge weights quantify a notion of dissimilarity where smaller values correspond to stronger interactions and the adjacency matrix elements are $d_{ij}\in[0,\infty)$, with $d_{ij}=\infty$ representing that $x_i, x_j \in X$ are not connected.
Note; however, that $d_{ij}$ is not strictly a distance for it may not obey a triangular inequality and can be asymmetric ($d_{ji}\neq d_{ij}$).
In fact, this lenient distance interpretation is what allows us to find distance backbone subgraphs for directed \cite{costa2023distance} and undirected \cite{simas2021distance} networks.
Both \textit{proximity adjacency matrix} elements, $p_{ij}$, and \textit{distance adjacency matrix} elements, $d_{ij}$, can represent the same network $G$ via to a continuous strictly decreasing map, $\varphi$, obeying the respective boundaries, $\varphi(0)=+\infty$ and $\varphi(1)=0$, which makes both quantities isomorphic \cite{simas2015distance}.
A canonical example is
\begin{equation}
    d_{ij} = \varphi(p_{ij}) = \frac{1}{p_{ij}} - 1,
    \label{eq:isomorphism}
\end{equation}
which we use for the remainder of the manuscript as the edge weights of $G$, unless explicitly noted otherwise.

Identifying the distance backbone subgraph of a complex networks requires determining if there is an alternative shortest-path for all existing edges \cite{simas2021distance, costa2023distance}.
Given any pair of nodes $x_i, x_j \in X$, a path is an ordered set of non-repeated nodes in which the initial element is $x_i$, the final element is $x_j$ and for any two consecutive nodes, there exists an edge connecting them. 
The path-length $\ell_{ij}$ for a given path $\Gamma = (x_i, x_{k_1}, \dots, x_{k_n}, x_j)$ can be given by a sum of the distance weights over the path's edges, which is traditionally done as
\begin{equation}
    \ell_{ij}(\Gamma) = d_{ik_1} + \dots + d_{k_nj}.
\label{eq:path_length_sum}
\end{equation}
However, as was stated in \cite{simas2015distance, simas2021distance}, there are infinite other ways of computing a path-length that don't necessarily rely on the summation of the edge distances.
These different ways were first presented in \cite{simas2015distance} and are called \textit{triangular distance norms} or \textit{td-norms} for short.
Formally, a \textit{td-norm} is a binary operation $g:[0,+\infty]\times[0,+\infty]$ that has a neutral element: $g(a, 0) = a$; is associative: $g(a, g(b, c)) = g(g(a, b), c)$; commutative: $g(a, b) = g(b, a)$ and monotonous: $a < b \Rightarrow g(a, c) < g(b, c)$, for all $a,b,c \in [0,+\infty]$.
Conceptually those axioms mean that a path-length measure should i) not depend on the order by which the edges are traversed (associativity and commutativity), ii) preserve the relative length order between paths passing via similar edge set (monotonicity), and iii) not be altered by a distance length of $0$ (neutral element).
With this in mind we can re-write the traditional path-length formula of Eq.~\ref{eq:path_length_sum} as:
\begin{equation}
\ell_{ij}(\Gamma) = g(d_{ik_1}, \dots, d_{k_nj}), 
\end{equation}
note that due to the associativity and commutativity of \textit{td-norms} we can extend the notation to accept $n$-ary inputs instead of just binary.
The traditional \textit{sum} $g(a,b) = a+b$ obeys all the axioms of a \textit{td-norm} as well as other notable examples like the \textit{euclidean sum} $g(a,b) = (a^2+b^2)^{\frac{1}{2}}$, the \textit{distance product} $g(a,b) = (a+1)(b-1) - 1$ or even the maximum function $g(a,b) = \max(a,b)$.
A given \textit{td-norm} defines how the length of paths are computed, leading to the shortest-path length between pair of nodes $x_i, x_j \in X$ being
\begin{equation}
d^{T,g}_{ij} = \underset{\Gamma}{\min} \: \ell_{ij}(\Gamma),
\end{equation}
where $T$ superscript refers to the transitivity (not transpose) induced in the shortest-path computation since the \textit{all-pairs shortest paths} (APSP) problem results in the transitive distance closure of the original graph \cite{simas2015distance}.
This means that $d^{T,g}_{ij} \leq g(d_{ik}, d_{kj}) \forall x_k \in X$, which allow us to define a neighborhood for every $x_i \in X$ leading to a topological space \cite{hausdorff1962set} induced by the path length measure $g$ \cite{simas2015distance}.
This space is composed of the edges satisfying $d_{ij} = d^{T,g}_{ij}$, for they are the ones obeying the generalized triangle inequality imposed by the \textit{td-norm} $g$, i.e. 
\begin{equation}
d_{ij} \leq g( d_{ik_1}, \dots, d_{k_nj}) \:\: \forall \: (x_i, x_{k_1}, \dots, x_{k_n}, x_j) \in \Gamma.
\label{eq:gen_trig_ineq}
\end{equation}
Those edges comprise the \textit{distance backbone} of a graph, $B^g \subset G$, associated with the \textit{td-norm} $g$ \cite{simas2021distance, costa2023distance}.
Furthermore, one can quantify how much each edges in $G$ breaks the generalized triangular inequality of Eq.~\ref{eq:gen_trig_ineq} by the \textit{semi-triangular distortion},
\begin{equation}
s_{ij}^g = \frac{d_{ij}}{d^{T,g}_{ij}}.
\label{eq:distortion_general}
\end{equation}
The edges in the backbone $B^g$ satisfy $s_{ij}^g=1$ and are said to be triangular, whereas the remaining edges are called semi-triangular and have $s_{ij}^g>1$ \cite{simas2021distance, costa2023distance}.

A particular case of \textit{distance backbones} obtained using $g \equiv +$ is the \textit{metric backbone}, denoted by $B^{m}$. In this case, we say that the edges in the backbone are metric, instead of triangular, while those with semi-metric distortion $s^m_{ij}>1$ are deemed semi-metric.
In previous work the metric backbone has been shown to maintain community structure and preserve SI epidemic spreading \cite{correia2023contact}. 
Also, it has been demonstrated that the semi-metric distortion identifies influential edges in spreading dynamics since the removal of edges with higher $s^m_{ij}$ have negligible effects on the transmission timescale and order of infection \cite{soriano2025quantifying}.
In other words, a complex network can be progressively sparsified by the decreasing order of semi-metric distortion while preserving dynamical features.

The limiting cases of distance backbones are the \textit{ultra-metric backbone} $B^{um}$, found by setting the path length measure to the \textit{maximum td-norm}, $g = g_{\text{max}}$ (Eq.~\ref{eq:max_tdnorm}), and the \textit{drastic backbone} $B^{\text{drastic}}$, found by setting the path length measure to the \textit{drastic td-norm}, $g = g_{\text{drastic}}$ (Eq.~\ref{eq:drastic_tdnorm}). 
The former is the smallest possible backbone and comprises the union of all minimum spanning forests \cite{rozum2024ultrametric}, while the latter is the unaltered original graph $G$ \cite{simas2021distance}.
\noindent\begin{minipage}{.5\linewidth}
\begin{equation}
  g_{\text{max}}(a, b) = \max(a, b)
  \label{eq:max_tdnorm}
\end{equation}
\end{minipage}\begin{minipage}{.5\linewidth}
\begin{equation}
  g_{\text{drastic}}(a, b) \equiv \begin{cases}
        a &\text{ if } b=0\\
        b &\text{ if } a=0\\
        +\infty &\text{ otherwise}\\
    \end{cases}
    \label{eq:drastic_tdnorm}
\end{equation}
\end{minipage}
Those limiting cases arise because all \textit{td-norms}, $g$, are bounded by $g_{\text{max}}$ and $g_{\text{drastic}}$ everywhere on $[0,+\infty]\times[0,+\infty]$ \cite{simas2015distance, simas2021distance}.
This means that, generally we have
\begin{equation}
 G = B^{\text{drastic}} = B^{g_{\text{drastic}}} \supseteq B^{g} \supseteq B^{g_{\max}} = B^{\text{um}}.
\end{equation}
Therefore, parameterizing $g$ such that it sweeps the space of \textit{td-norms} from $g_{\text{max}}$ to $g_{\text{drastic}}$ will give us a sequence of distance backbones from the smallest subgraph to the original network.
In this work we explore the consequences of such parametrization in inferring structural and dynamical network properties. 

\section{Results}

\subsection{Establishing the Distance Backbone Synthesis}
\label{sec:dbs_intro}

\begin{figure}[t!]
\centering
\includegraphics[width=1\linewidth]{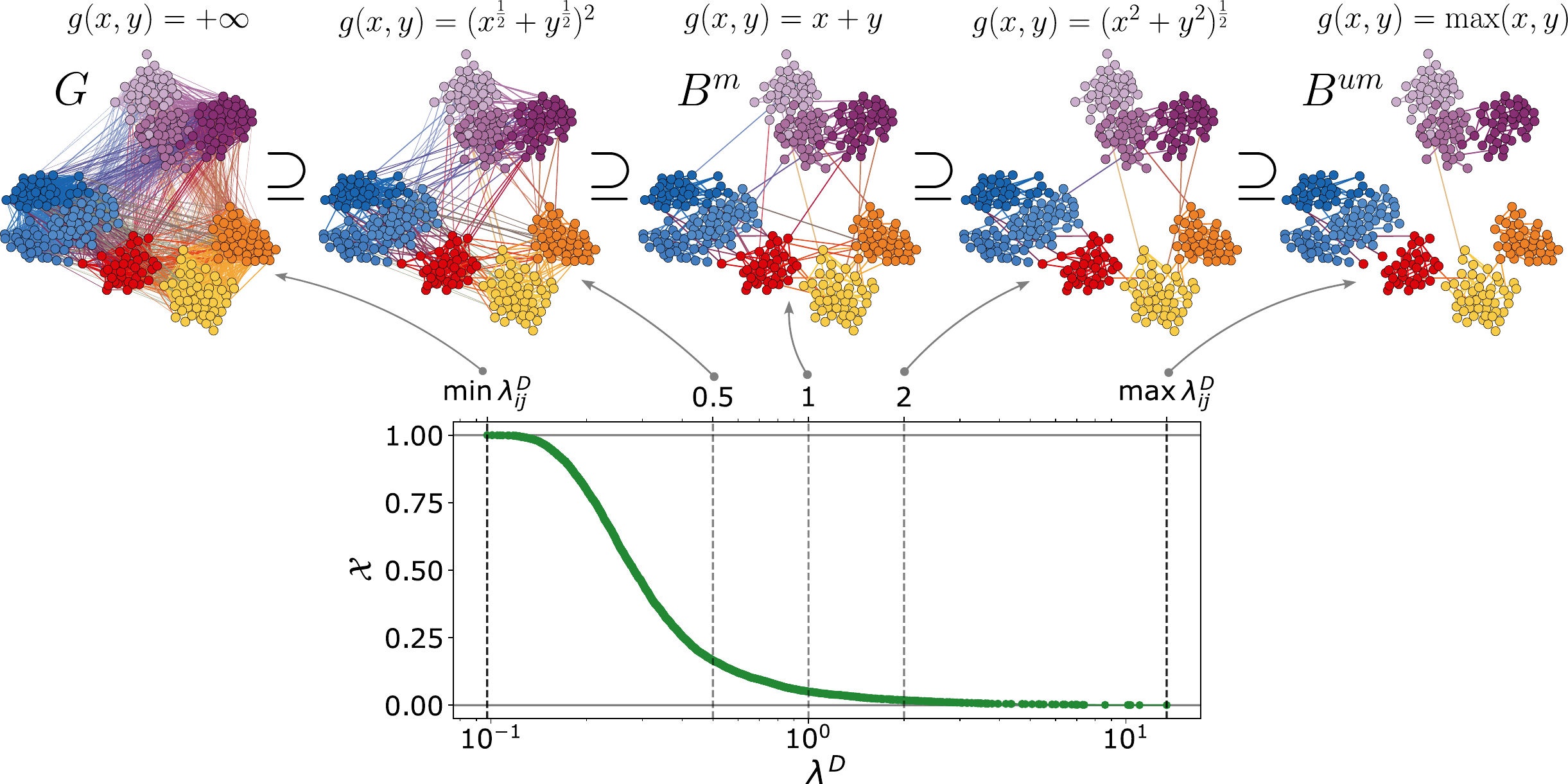}
\caption{Progressive sparsification of a social contact network from a High School in Marseille, France \cite{mastrandrea2015contact} described in Sec.~\ref{sec:social_networks} by the Distance Backbone Synthesis, $\textbf{DBS}$. (\textit{Below}) The relative size, $\mathcal{X}$ of the subgraph found via the backbone synthesis based on path-length measure in Equation~\ref{eq:dombi_tdnorms} varying with $\lambda^D$. When $\mathcal{X}=1$ there is no sparsification and when $\mathcal{X}=0$ we have the largest possible sparsification. (\textit{Above}) Important subgraphs in this sparsification: the drastic backbone $B^{g_{0}}$, which is the original graph $G$, the $B^{g_{0.5}}$ backbone, the metric backbone $B^{m} = B^{g_{1}}$, the euclidean backbone $B^{g_{2}}$ and the ultra-metric backbone $B^{um} = B^{g_{+\infty}}$.}
\label{fig:dombi_sparsification}
\end{figure}

In order to sweep the different ways in which a path length can be measured in complex networks \cite{simas2015distance} we introduce a novel sparsification method called the Distance Backbone Synthesis ($\textbf{DBS}$), which is a systematic way of computing all distance backbones within a range of parameterized triangular distance norms, $g_{\lambda}$.
Without loss of generality we can define a $g_\lambda$ such that i) for $\lambda_1 < \lambda_2$ we have $g_{\lambda_1} > g_{\lambda_2}$ in its entire domain, $[0, +\infty]\times[0, +\infty]$, and ii) is bounded by $g_{\text{drastic}}$ and $g_{\text{max}}$.
Since for an edge from $x_i$ to $x_j$ to be in the backbone $B^{g_{\lambda_2}}$ it must obey $d_{ij} \leq g_{\lambda_2} (d_{ik}, d_{kj}) \forall x_k\in X$, then it follows that $d_{ij} \leq g_{\lambda_1} (d_{ik}, d_{kj}) \forall x_k\in X$, implying that the same edge is also in the backbone $B^{g_{\lambda_1}}$.
In fact, the backbones found via this ordered $g_\lambda$ are all found to be encapsulated in each-other, forming a chain
\begin{equation}
B^{\text{drastic}} = B^{g_0} \supseteq \dots \supseteq B^{g_{\lambda_1}} \supseteq \dots \supseteq B^{g_{\lambda_2}} \supseteq \dots \supseteq B^{g_{+\infty}} = B^{\text{um}}.
\end{equation}
This newly found encapsulation of backbones, ranging from the smallest backbone possible to the whole network, implies that for each edge in the original graph, there is a first $\lambda_{ij}$ for which that edge becomes part of the corresponding backbone.
Moreover, since $\lambda$ is generally a continuous parameter and there is a large heterogeneity of distance weights in real-world networks, each edge has a unique $\lambda_{ij}$.
This induces an order of the edges, which acts as a synthesis of all the possible distance backbones from the family of path-length measures $g_\lambda$.
Also, since $g_{\lambda}$ enforces a topology via its shortest-path closure, then finding $\lambda_{ij}$ for each edge explains all the sufficient shortest-paths of a network in every topological space it covers.
In Supplementary Section B 
we show options for $g_\lambda$ that are in agreement with the above arguments, while focusing on a particular case in the manuscript.

Consider the cases where
\begin{equation}
    g_\lambda(x,y) = (x^\lambda + y^\lambda)^{\frac{1}{\lambda}},
    \label{eq:dombi_tdnorms}
\end{equation}
which is a \textit{td-norm} family corresponding to the family of Dombi triangular norms (\textit{t-norms}) in fuzzy logic \cite{klir1995fuzzy, simas2015distance}, we have a continuous parameter $\lambda \in [0,\infty]$ such that i) when $\lambda \rightarrow +\infty$ the \textit{td-norm} converges to $g \equiv \max$, whereas when $\lambda \rightarrow 0$ they converge to $g \equiv g_{drastic}$, and ii) $g_{\lambda_1} > g_{\lambda_2}$ for $\lambda_1 < \lambda_2$ in the domain $[0, +\infty]\times[0, +\infty]$.
An advantage of this parametrization, beyond its algebraic simplicity, is the fact that it generalizes the usual sum, as $g_1 (x,y) = x + y$, which is the canonical operation used in the computation of \textit{shortest-paths}.
It also generalizes pathfinder networks \cite{schvaneveldt1990pathfinder}, which are often used in the analysis of high-dimensional numerical datasets in the context of machine learning and clustering algorithms \cite{mailagaha2022generalized, lange2014applications}, to go beyond the Minkowski $r$-metrics (or $L^r$-norms), with $r\in [1, +\infty]$, and allow any positive real values, $\lambda \in [0,\infty]$.
Another choice of $g_\lambda$ is discussed in Supplementary Section B.

The Distance Backbone Synthesis ($\textbf{DBS}$) allow us to sweep through different levels of sparsification by varying a threshold below which an edge with $\lambda_{ij} < \lambda^D$ (the superscript $D$ denotes that the value of $\lambda$ is found using Equation~\ref{eq:dombi_tdnorms} from the family of Dombi \textit{t-norms}) is in the sparsified subgraph (Fig.~\ref{fig:dombi_sparsification}).
Note that the adjusted relative size of the subgraph $\mathcal{X}$, which ranges from $\mathcal{X}=0$ for the ultrametric backbone to $\mathcal{X}=1$ for the entire graph, is the complementary cumulative distribution function of $\lambda^D$ and its sigmoidal shape in semi-log scale suggest a log-normally distributed $\lambda_{ij}$.
We perform the curve fit of $\mathcal{X}(\lambda^D)$ for the empirical social networks described in Sec.~\ref{sec:social_networks} and found excellent agreement with log-normal distribution, lowest $R^2$ of 0.98, as shown in Supplementary Table 2.
The average mode of this log-normal distributions is $0.3\pm 0.1$, meaning not only that most edges in those networks lie in some sublinear/fractal topological space ($\lambda_{ij} \simeq 0.3$), which is expected considering that their metric backbone is much less than 30\% of the network, but also points that $\lambda^D\approx0.3$ suffices to considerably sparsify many social contact networks.

$\textbf{DBS}$ brings a new parameter of edge importance based on distance backbones.
Other sparsification methods also characterize edge importance in different ways.
Most popularly, edge importance is determined in weight thresholding (\textbf{W}) by the interaction strength and in the disparity filter (\textbf{DispFilt}) \cite{serrano2009extracting} by the statistical significance relative to a null model.
While dynamically relevant edges have been found by assigning their importance to its weighted effective resistance (\textbf{wEffRes}) \cite{mercier2022effective} and semi-metric distortion ($\textbf{SmD}$) \cite{soriano2025quantifying}.
In what follows, we will compare how those different sparsification methods preserve structural and dynamical information from the original network.

\subsection{Distance Backbones Preserves Centrality and Dynamics}
\label{sec:methods_comparison_results}

\begin{figure}[t!]
\centering
\includegraphics[width=\linewidth]{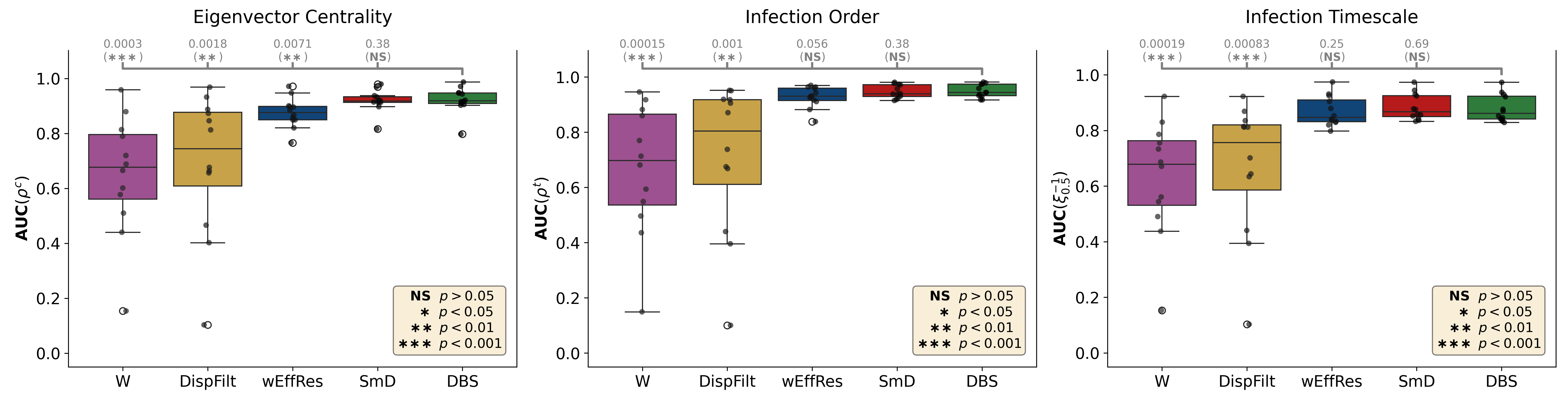}
\caption{\textbf{Sparsification effect on social contact networks.} 
From the 12 social contact networks described in Sec.~\ref{sec:social_networks}, we have a distribution of the area under the curve for different sparsification levels of (\textit{Left}) eigenvector centrality rank, $\rho^C$, (\textit{Middle}) average time of infection, $\rho^t$, and (\textit{Right}
) time at which half of the nodes becomes infected in the original network relative to the same time measured in the sparsified network, $\xi^{-1}_{0.5}$ (Details in Sec.~\ref{sec:performance_metrics}). 
Mann-Whitney U-Test was used to establish if the distribution for $\textbf{DBS}$ (green) is significantly larger than the distribution for each of the sparsifcation methods discussed in text (see legend), obtaining the annotated $p_{vals}$.}
\label{fig:all_auc_networks}
\end{figure}

One of the most widely used measure of node importance is the eigenvector centrality \cite{bovet2021centralities}, because it account for the quantity and quality of the pathways in which a node is involved.
Therefore, by constructions, one expects that network sparsification methods will modify the absolute centrality of nodes upon them losing edges.
Nevertheless, in practice, the ranks of eigenvector centrality in a network is a more appropriate marker to determine node importance in various dynamical processes \cite{jackson2008social, pastor2015epidemic, fletcher2018structure}, due to its absolute variability.
Thus, we are interesting in identifying which sparser version of the original network best preserves the original centrality rank of the nodes, $\rho^C$.

Eigenvector centrality ranking, $\rho^C$, is best preserved by sparsification methods removing shortest-path redundancy.
We summarize this finding for all 12 social contact networks analyzed by measuring the area under the curve, $\textit{AUC}$, of $\rho^C$ until a maximum sparsification, $\Tilde{\mathcal{X}}$, as described in Sec.~\ref{sec:auc_summary_methods}.
Each of those empirical networks correspond to a dot in the boxplots of Fig. \ref{fig:all_auc_networks} for each sparsification method.
Overall, the distribution of $\textit{AUC}(\rho^{C})$ is statistically higher when edge importance is measured by $\textbf{DBS}$ or $\textbf{SmD}$ than in comparison with \textbf{wEffRes}, \textbf{DispFilt}, and \textbf{W} as shown in Fig. \ref{fig:all_auc_networks} (\textit{Left}).
Since $\textit{AUC}(\rho^{C})$ is bounded from above by $1$, this result shows that removing shortest-path redundancy, either via $\textbf{SmD}$ or $\textbf{DBS}$, is a more generally relevant methodology to sparsify complex networks while preserving eigenvector centrality rank.
Furthermore, in Supplementary Fig. 9
, we show that $\textbf{DBS}$ can find sparser connected subgraphs than \textbf{wEffRes}, \textbf{DispFilt}, and \textbf{W} for the same desired $\rho^C$ in those networks, while it is only able to surpass $\textbf{SmD}$ performance for empirical networks with more random connectivity (Supplementary Fig. 13).

Network sparsification based on shortest path redundancy preserves spreading dynamics at a microscopic scale.
This is quantified by the order at which each nodes gets infected, $\rho^t$, in an epidemic spreading agent-based simulation of a Susceptible-Infected (SI) model as described in Sec.~\ref{sec:spreading_methods}.
Fig.~\ref{fig:all_auc_networks} (\textit{Middle}) summarizes our findings for all 12 social contact networks analyzed via $\textit{AUC}(\rho^{t})$ until a maximum sparsification, $\Tilde{\mathcal{X}}$, as described in Sec~\ref{sec:auc_summary_methods}.
It is striking that a more informative method such as $\textbf{DispFilt}$ performs just as well as the simple $\textbf{W}$, while the other methods have higher performance albeit statistically similar.
Interestingly, the likelihood of $\textbf{DBS}$ being statistically larger than $\textbf{wEffRes}$ is not far from being significant ($p_{val}=0.056$), and the distribution of the first is less skewed and with a higher mean and median $\textit{AUC}(\rho^{t})$ than of the second.
More importantly, $\textbf{DBS}$ is a more traceable methodology for it does not change the edge weights in an irreversible operation as $\textbf{wEffRes}$ does (See Sec.~\ref{sec:sparsification_schemes}).

Network sparsification based on shortest path redundancy preserves the timescale of the SI spreading, $\xi^{-1}_{0.5}$.
Comparing all 12 social contact networks via the distribution of the area under the curve $\textit{AUC}(\xi^{-1}_{0.5})$ in Fig.~\ref{fig:all_auc_networks} (\textit{Right}), $\textbf{wEffRes}$, $\textbf{SmD}$, and $\textbf{DBS}$ have a superior performance than $\textbf{DispFilt}$ and $\textbf{W}$.
The high performance of $\textbf{wEffRes}$ in not surprising since it involves approximating the network Laplacian, which encodes a diffusion process in the network and as such is expected to capture well the spreading timescale.
However, the performance of $\textbf{SmD}$ and $\textbf{DBS}$ are both statistically comparable to \textbf{wEffRes} while they produce higher mean and median $\textit{AUC}(\xi^{-1}_{0.5})$ with the added feature of keeping the meaning of edge weights.
Both $\textbf{SmD}$, and $\textbf{DBS}$ performed overwhelmingly better than other methods when we considered networks with more random connectivity patterns (See Supplementary Sec. C.7).
Meanwhile, \textbf{W} and \textbf{DispFilt} performed extremely poorly because they removed bridges very early on in the sparsification process, making the networks disconnected and hindering their spreading ability.
On the other hand, the variability in the performance of $\textbf{wEffRes}$ implies that it less generalizable then $\textbf{SmD}$ and $\textbf{DBS}$.
Therefore, this higher performance highlights the wide applicability of distance backbones in diverse contexts and with distinct types of networks.

\subsection{DBS Reveals Optimal Backbones}

\begin{figure}[t!]
\centering
\includegraphics[width=\linewidth]{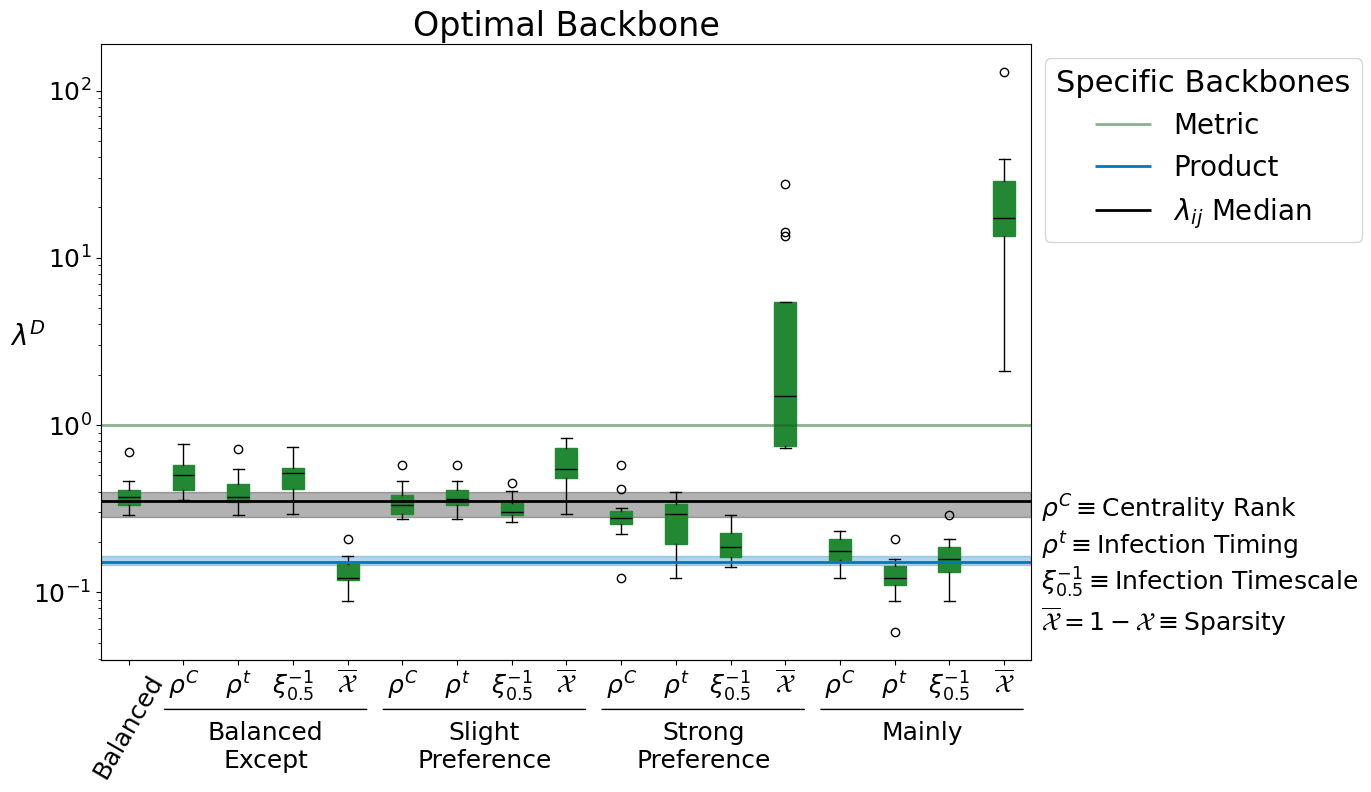}
\caption{\textbf{Optimal backbone that preserve desired network features.}
We measure the optimal $\lambda^D$ found via a weighted sum of the eigenvector centrality ranking, $\rho^C$, infection timing ranking, $\rho^t$, infection timescale, $\xi^{-1}_{0.5}$, and sparsity, $1-\mathcal{X}$, over various scenarios described in Section~\ref{sec:optimization_methods} per network.
We plot the distribution of the optimal $\lambda^D$ over the 12 social contact networks considered.
The particular cases of the metric backbone (light green) and the product backbone (light blue) are highlighted.
The later is a small range of values for it can only be defined in a different parametrization, as discussed in Supplementary Section B.
We also shown the median and inter-quartile range, over the 12 networks, of the median $\lambda_{ij}$ in black.
}
\label{fig:multi_objective}
\end{figure}

Having established the $\textbf{DBS}$ as one of the best network sparsification methods capable of optimizing preservation of network structure and dynamics, we investigate if there is an overall distance backbone capable of preserving desired network features.
After all, by measuring the length of paths in a network using Equation~\ref{eq:dombi_tdnorms} with $\lambda=\lambda^D$ results in the backbone subgraph $B^{g_{\lambda}} \subseteq G$.
Fig.~\ref{fig:multi_objective} shows the distribution of $\lambda^D$ found from optimizing the weighted sum of the eigenvector centrality ranking, $\rho^C$, infection timing ranking, $\rho^t$, infection timescale, $\xi^{-1}_{0.5}$, and sparsity, $1-\mathcal{X}$, over various scenarios (see Methods Section~\ref{sec:optimization_methods}).
Our results show that an optimal backbone would be between the product backbone, found by the path length measure $g(x, y) = (x+1)(y+1) - 1$, and the metric backbone, found by the path length measure $g(x, y) = x+y$.
Despite neither being an overall optimal, those cases give us the bounds for a lenient and stringent sparsification to preserve network features.
A noteworthy exception is that the product backbone can perform very well in preserving infection timing rank, $\rho^t$, and timescale, $\xi^{-1}_{0.5}$, in many social networks.

The search for an optimal backbone also leads us to the finding that most of the connections can be removed in order to preserve multiple network features.
As mentioned in Section~\ref{sec:dbs_intro}, most of the edges in all social networks considered have $\lambda_{ij} \lesssim 0.3$.
Independently, Fig.~\ref{fig:multi_objective} shows it to be a higher lower bound than the product backbone ($\lambda^D \approx 0.15$) for an optimal backbone where the weighted sum optimization coefficients are closer to one another (see Methods Section~\ref{sec:optimization_methods} for detailed values).
Therefore, setting $\lambda^D \approx 0.3$ allows one to achieve large amounts of sparsification while still preserving at least two of eigenvector centrality, $\rho^C$, infection timing rank, $\rho^t$, and infection timescale, $\xi^{-1}_{0.5}$.
Nevertheless, all three performance metrics are above 80\% of their original value for $\lambda^D \approx 0.3$ (Supplementary Figure 16).

Each optimization scenario has a characteristic topological space identified by the $\textbf{DBS}$ parameter $\lambda^D$.
However, due to the computational costs in measuring $\lambda_{ij}$, we recommend users of this methodology to choose \textit{a priori} the desired topological space depending on the sparsification leniency.
Specific $\lambda^D$ for a given percent threshold of the performance metrics can be found in Supplementary Figure 16.
A more computationally efficient alternative is the $\textbf{SMDS}$ \cite{soriano2025quantifying}, in which a multi-objective optimization (see Supplementary Fig. 15)
finds that a lenient sparsification is to remove edges with semi-metric distortion $s_{ij}^m > 50$, while those with $2 < s_{ij}^m < 50$ are needed to satisfy multiple objectives.
The main difference in the optimal subgraph for both $\textbf{DBS}$ and $\textbf{SMDS}$ is in the spread of their distributions, where the $\textbf{DBS}$ provide a more precise threshold for an optimal subgraph across networks.

\section{Discussion}
\label{sec:discussion}

In this study we investigated how distance backbones can be used to create sparser representations of complex networks which are dynamically and structurally faithful.
In particular, by introducing the Distance Backbone Synthesis ($\textbf{DBS}$) we were able to identify which topological space each edge belongs to, based on how the length of paths are measured (Equation~\ref{eq:dombi_tdnorms}). 
This identification is given by the edges in the distance backbone for a given synthesis parameter, $\lambda^D$, whose distance are compared to the length of indirect paths.
Thus, the different ways to compute path lengths correspond to attributing different costs to indirect paths.
Larger values of $\lambda^D$ correspond to less costly indirect paths, which induces larger sparsification for it results in smaller distance backbones.

Network sparsification facilitates the analysis and simulation of social, biological and many other complex systems \cite{soriano2025quantifying}.
However, this reduced representation should not interfere with a system function.
As a proxy for how well sparsified network representation preserves the necessary information in a systems we evaluate the ability of state-of-the-art sparsification methods to preserve eigenvector centrality ranking and SI spreading (Section~\ref{sec:methods_comparison_results}).
In those tasks, $\textbf{DBS}$ allows for a progressive network sparsification which outperforms more traditional methods \cite{yan2018weight, serrano2009extracting, mercier2022effective} across many networks with diverse connectivity patterns, and has a comparable performance with the semi-metric distortion sparsification ($\textbf{SmD}$) \cite{soriano2025quantifying}.
Despite comparable performances, $\textbf{DBS}$ brings the added explainability about which topological space each edge belongs to.
Meanwhile, $\textbf{SmD}$ measures how far from the topological space defined by $\lambda^D = 1$ an edge is, via the semi-metric distortion.
Therefore, the agreement in performance between those methods is because they look at shortest-path redundancy \cite{simas2021distance} in comparable ways.

Despite the performance gains and explainability of the Distance Backbone Synthesis, it has some drawbacks.
Currently, measuring the edge importance $\lambda_{ij}$ is done via an exhaustive search, which is computationally expensive.
Moreover, similarly to other progressive sparsification methods, $\textbf{DBS}$ requires practitioner decision on which threshold $\lambda^D$ to choose from for their network analysis.
Our multi-objective optimization has uncovered a lenient and strict bounds as the non-parametric product and metric backbones \cite{simas2021distance}, respectively, with the latter being already proven \cite{correia2023contact} to best preserve SI dynamics.
Moreover, since the optimal backbone for a balanced optimization corresponds to $\lambda^D \approx 0.3$, this points to the $\textbf{DBS}$ parameter being a characteristic of the fractal topology required for spreading dynamics \cite{grassberger1986spreading}.
This opens up future work addressing the extent to which more complex dynamical models such as Susceptible-Infected-Susceptible and Susceptible-Infected-Recovered epidemics, and synchronization phenomena require fractal topology.

Another noteworthy aspect of the analysis performed in this work is the large diversity in the networks studied, which have very different sizes, connectivity and densities.
Despite still not being completely understood how each of those network properties affects sparsification methods, our results comparing the performance distributions show that removing shortest-paths redundancy is a simple way to sparsify networks without losing important information.

In the main text, we have focused on the distance backbone synthesis under the Dombi path length family (Eq.~\ref{eq:dombi_tdnorms}), which generalizes pathfinder networks \cite{schvaneveldt1990pathfinder}.
It is important to note that the $\textbf{DBS}$ could be performed using another family of path length measures, as done in Supplementary Section B.
Furthermore, for allowing any positive real values of $\lambda_{ij}$ we are bound to have a higher resolution on edge importance, which our results shown to be relevant for inferences on networks.
One could even go beyond distance backbones, to its parent theory of distance closures \cite{simas2015distance}, and possibly identify other subgraphs that are capable of preserving a non-shortest-path distance closure.
Defining such subgraph and identifying its relevance is left as future work.

\section{Materials and Methods}

In this section we describe how the empirical social networks used in our study are built, particularly the strength of interactions in them. 
Then, we explain the network sparsification methods considered in this study. 
Finally, we present the definitions of the metrics used to quantify the sparsification performances on node centrality and spreading dynamics.

\subsection{Social Network Construction}
\label{sec:social_networks}

We construct empirically measured social contact networks publicly available in previous studies (cited in Tab.~\ref{tab:network-stats}), where each node represents a person and different people are connected based on the frequency of their physical proximity. 
Using radio frequency identifications (RFIDs) and Bluetooth devices, which have accompanied each individual during a certain amount of time, it was registered who was in close physical proximity with whom during short intervals of about $20$ seconds.
Denoting by $r_{ij}$ the number of intervals between person $x_i$ and $x_j$ during the overall data collection, and $r_{ii}$ the total number of close interactions that the person $x_i$ had, the strength of their interaction is found by
\begin{equation}
p_{ij} = \frac{r_{ij}}{r_{ii}+r_{jj}-r_{ij}}.
\end{equation}
This value can be regarded as a similarity measure between two people, which has been called a \textit{proximity measure} in other studies \cite{simas2015distance, simas2021distance, costa2023distance}, a convention we will follow.
That is, a graph $G$ has a set of nodes $X$ and weighted edges $p_{ij}\in[0,1]$ which is regarded as the intensities by which two nodes $x_i, x_j \in X$ are connected.
From $p_{ij}$ we measure the isomorphic interaction distance $d_{ij}$ between nodes using equation~\ref{eq:isomorphism}.

\begin{table}[h!]
    \centering
    \caption{Characteristics of the networks used in this work.}
    \begin{tabular}{>{\bfseries}l|c|c|c|c|c}
        \hline
        \textbf{Network} & \textbf{Nodes} & \textbf{Edges} & \textbf{Density} & $\mathbf{\mathcal{X}(\lambda^D=1)\%}$ & \textbf{Source} \\
        \hline
        \texttt{kenya-households}      & 31   & 216   & 0.465 &  13.98 & \cite{kiti2016quantifying} \\ \hline
        \texttt{hospital}              & 75   & 1139  & 0.410 & 13.43 & \cite{vanhems2013estimating} \\ \hline
        \texttt{malawi-village}        & 84   & 346   & 0.099 & 9.51 & \cite{ozella2021using} \\ \hline
        \texttt{workplace1}            & 92   & 755   & 0.180 & 20.63 & \cite{genois2015data} \\ \hline
        \texttt{conference}            & 113  & 2196  & 0.347 & 9.40 & \cite{isella2011whats} \\ \hline
        \texttt{workplace2}            & 217  & 4274  & 0.182 & 13.04 & \cite{genois2018can} \\ \hline
        \texttt{french-primary-school} & 242  & 8317  & 0.285 & 6.80 & \cite{stehle2011high} \\ \hline
        \texttt{french-high-school}    & 327  & 5818  & 0.109 & 5.04 & \cite{mastrandrea2015contact} \\ \hline
        \texttt{us-elementary-school}  & 339  & 16546 & 0.289 & 4.87 & \cite{toth2015role} \\ \hline
        \texttt{us-middle-school}      & 591  & 56867 & 0.326 & 5.21 & \cite{toth2015role} \\ \hline
        \texttt{copenhagen-university} & 692  & 79530 & 0.333 & 4.72 & \cite{sapiezynski2019interaction} \\ \hline
        \texttt{us-high-school}        & 788  & 118291& 0.381 & 7.22 & \cite{salathe2010high} \\ 
        \hline
    \end{tabular}
    \label{tab:network-stats}
\end{table}

Our approach is easily generalizable to networks beyond the social domain as suggested in \cite{soriano2025quantifying}.
The diversity of social contact networks samples we have in Tab.~\ref{tab:network-stats}, crossing multiple orders of magnitude in both numbers of nodes and edges, hints to the generalizability of our findings.
That is, weighted graph in which the nodes are connected by a measure of similarity (or dissimilarity) have edges belonging to different topological spaces, determined by the Distance Backbone Synthesis parameter $\lambda$, which our results show to be sufficient for dynamical inference.

\subsection{Sparsification methods}
\label{sec:sparsification_schemes}

Many network sparsification methods define a parameter quantifying the importance of each edge in a network \cite{serrano2009extracting, mercier2022effective, soriano2025quantifying}. 
This quantification induces an increasing (decreasing) sparsification order whereas the least important edges are those with smaller (larger) parameter values and could be removed.
In this work, we propose the Distance Backbone Synthesis (\textbf{DBS}), which can progressively sparsify networks by removing edges in increasing value of $\lambda_{ij}$.
\textbf{DBS} performance is compared with other sparsification methods described below and summarized in Tab.~\ref{tab:methods-used}.

\begin{table}[h!]
    \centering
    \renewcommand{\arraystretch}{1.2}
    \caption{Sparsification Methods Used}
    \begin{tabular}{c|c|c|c|c}
        \hline
        \textbf{Label} & \textbf{Name} & \textbf{Parameter} & \textbf{Sparsification Order} & \textbf{Relevant Sources} \\
        \hline\hline
        \textbf{W}        & Weight Thresholding           & $p_{ij}$                      & increasing ($\nearrow$) & \cite{yan2018weight} \\ \hline
        \textbf{DispFilt} & Disparity Filter              & $\alpha_{ij}$                 & decreasing ($\searrow$) & \cite{serrano2009extracting} \\ \hline
        \textbf{wEffRes}  & Weighted Effective Resistance & $p_{ij}R^{e}_{ij}$            & increasing ($\nearrow$) & \cite{mercier2022effective, spielman2011graph} \\ \hline
        \textbf{SmD}      & Semi-Metric Distortion        & $s^m_{ij}$                    & decreasing ($\searrow$) & \cite{soriano2025quantifying} \\ \hline
        \textbf{DBS}  & Distance Backbones Synthesis     & $\lambda_{ij}$     & increasing ($\nearrow$) & - \\ \hline
    \end{tabular}
    \label{tab:methods-used}
\end{table}

The most simplistic way of sparsifying a network is the removal of edges which encode weak interactions, this is the case of the weight thresholding (\textbf{W}), where edges are removed by increasing order of the proximity weight $p_{ij}$, thus lower proximities are removed first for they represent weaker interactions.

A more informed approach, the disparity filter (\textbf{DispFilt}) \cite{serrano2009extracting}, assumes a null model for the network connectivity in order to quantify an edge statistical significance, $\alpha_{ij}$, which is used to remove edges in decreasing order, since edges with lower $\alpha_{ij}$ values are statistically more likely to be relevant.
More explicitly, \textbf{DispFilt} consider a null-model in which the probability density function of a normalized edge weight from a node $x_i$ taking value $p\in[0,1]$ is given by $ F_{k_i}(p) = \int^p_0 ({k_i}-1)(1-x)^{{k_i}-2} \dd x$, where ${k_i}$ is the node degree, leading to $\alpha_{ij} = 1 - \min(F_{k_i}(p_{ij}), F_{k_j}(p_{ij}))$ for an undirected network.
In other words, an edge is kept if it is significant for at least one of the nodes it connects.

Another methodology, which draws inspiration from electrical networks, is the weighted effective resistance (\textbf{wEffRes}) \cite{mercier2022effective} which removes edges in increasing order of a measure of weighted resistance between node pairs, $p^R_{ij}$, since smaller values were proposed to be less relevant for global information exchange.
The effective resistance between any two nodes has information about all the possible paths connecting them since $p^R_{ij}=p_{ij} R^{e}_{ij}$, where $R^{e}_{ij} = (e_i - e_j)^{\top} L^{\dag} (e_i - e_j)$ is the projection of the Moore-Penrose pseudo-inverse of the Laplacian matrix, $L^{\dag}$, unto the canonical basis $e_i,e_j$.
Usually, computing $L^{\dag}$ is often a very difficult task especially for very large networks; thus, some approximation algorithm is required \cite{mercier2022effective}.

Ultimately, we also compare \textbf{DBS} with another method that takes into account the metric topology of weighted graph, the semi-metric distortion sparsification ($\textbf{SmD}$) \cite{soriano2025quantifying}.
In this method, edges are removed in decreasing order of their semi-metric distortions, $s^{m}_{ij}$ since they have been shown to not be as important as those with $s^{m}_{ij}\gtrapprox1$ to spreading dynamics \cite{soriano2025quantifying}.
The semi-metric distortion of an edge is given by
\begin{equation*}
    s_{ij}^m = \frac{d_{ij}}{d^{T,m}_{ij}}\text{, where } d^{T,m}_{ij} = \underset{\Gamma}{\min} \: (d_{ik_1} + \dots + d_{k_nj}),
\end{equation*}
for a given path $\Gamma = (x_i, x_{k_1}, \dots, x_{k_n}, x_j)$.

It is important to note that methods such as $\textbf{W}$, $\textbf{DispFilt}$, and $\textbf{wEffRes}$ do not guarantee connectivity preservation but have been empirically found to maintain it up to a certain point.
On the other hand, methods based on the existence and preservation of shortest paths, like $\textbf{SmD}$ and $\textbf{DBS}$, always preserve connectivity as they ensure at least one path between any two nodes.
In the case of $\textbf{SmD}$, connectivity is guaranteed only up to the metric backbone size. At this point, all edges are indistinguishable under this method ($s^{m}_{ij}=1$), and removing any edge under these conditions might disconnect the network.
Conversely, $\textbf{DBS}$ preserves connectivity throughout the largest possible sparsification range, down to the union of all minimum spanning trees in the graph, also called the \textit{ultrametric backbone}, $B^{um}$ \cite{rozum2024ultrametric}. 
Additionally, in the networks used in this study, the size of the ultrametric backbone corresponds to the number of nodes minus one, indicating that these networks have a unique minimum spanning tree.

\subsection{Performance Metrics}
\label{sec:performance_metrics}

\begin{figure}[t!]
    \centering
    \includegraphics[width=\linewidth]{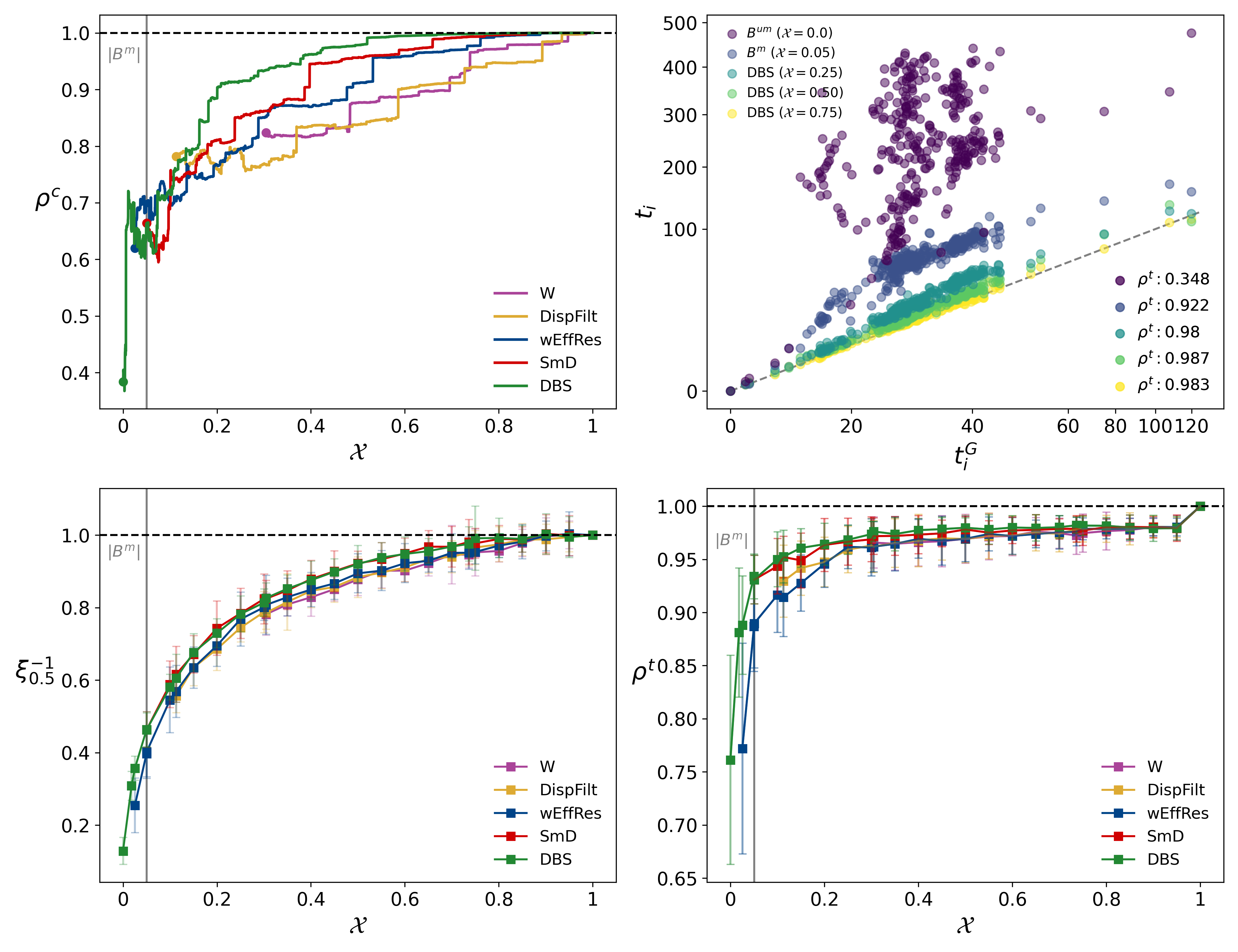}
    \caption{Example of the sparsification performance metrics on the social contact network from a high school in Marseilles, France, \cite{mastrandrea2015contact} described in Sec.~\ref{sec:social_networks}.
        Curves for each sparsifcation method (per legend) are shown for different levels of sparsification, $\mathcal{X}$. 
        (\textit{Top-Left}) Spearman's rank correlation between the eigenvector centrality of nodes in the sparsified network relative to the original network, $\rho^C$.
        (\textit{Bottom-Left}) Time at which half of the nodes becomes infected in the original network relative to the same time measured in the sparsified network, $\xi^{-1}_{0.5}$. Error-bars quantify the standard-deviation relative to choosing different seed nodes in the simulations.
        (\textit{Top-Right}) Average time in which each node got infected in the original graph considering a specific seed node, horizontal axis, and in networks sparsified by the Distance Backbone Synthesis $\textbf{DBS}$, vertical axis. Each color corresponds to a different sparsification level, $\mathcal{X}$, which is annotated with the Spearman's rank correlation between those times, $\rho^t$.
        (\textit{Bottom-Right}) Spearman's rank correlation between average time of infection, $\rho^t$. Error-bars quantify the standard-deviation relative to choosing different seed nodes in the simulations.
    }
    \label{fig:performance_metrics_example}
\end{figure}

In this subsection we defined the performance metrics used to compare the different sparsification methods.
They are measured at different adjusted relative size of the subgraph, $\mathcal{X}$, defined with respect to the sizes of the limiting distance backbones subgraphs: the \textit{ultra-metric} $B^{um}$ and the \textit{drastic} $B^{\text{drastic}}$.
That is, for a network $G$ containing $\abs{G}$ edges which is sparsified until the subgraph $G'$, then
\begin{equation}
    \mathcal{X} = \frac{\abs{G'}-\abs{B^{um}}}{\abs{B^{\text{drastic}}}-\abs{B^{um}}} = \frac{\abs{G'}-\abs{X}+1}{\abs{G}-\abs{X}+1},
\end{equation}
since, in the studied networks, $\abs{B^{\text{drastic}}} = \abs{G}$ and $\abs{B^{um}}=\abs{X}-1$, where $X$ is the set of nodes.
Therefore, $\mathcal{X}=0$ for the ultrametric backbone and $\mathcal{X}=1$ for the entire graph.

\subsubsection{Node Centrality}
\label{sec:eigvec_methods}

Several centrality measures have been proposed to quantify the relevance of each node in a network.
In this work we focus specifically on eigenvector centrality, which is among the most widely used measures of node centrality in complex networks \cite{bovet2021centralities}. 
This centrality measure is very popular because it considers the number of a nodes' connections as well as their quality, attributing high centrality scores to nodes connected to other high centrality nodes. 
Hence, each node's centrality, $c_i$, corresponds to its entry in the eigenvector associated with the largest eigenvalue, $\Lambda$, of the network proximity adjacency matrix with elements $p_{ij}\in[0, 1]$:
\begin{equation}
    c_i = \frac{1}{\Lambda} \sum_{x_j\in X} p_{ij} c_j.
\end{equation}
The exact value of $c_i$ is not particularly informative unless in comparison to centrality of the other nodes.
Therefore, in order to compare how well a sparsified graph preserves this topological network feature we compute the Spearman's rank correlation between the node eigenvector centralities in the original network and a sparsified subgraph, $\rho^C$, for different amounts of sparsification $\mathcal{X}$.
In the sparsified subgraphs, the proximity weights are the same for the edges kept, and set to zero for the edges removed in order to compute the new centrality.
Larger values of $\rho^C$ correspond to better preservation of centrality rank.

\subsubsection{Spreading Dynamics} 
\label{sec:spreading_methods}

In this work we focus on the Susceptible-Infected (SI) epidemiological model to simulate spreading dynamics on networks \cite{pastor2015epidemic}. 
In this model a single seed node starts in the Infected state, whereas the rest of the nodes start in the Susceptible state. 
Then, at each discrete timestep $t$, each infected node infects its neighbor nodes with probability $\beta p_{ij}$, where $p_{ij}$ corresponds to the proximity value of the edge connecting them and $\beta$ is global parameter that regulates the timescale of the spreading. 
Once all the nodes become infected, the spreading terminates and thus, the process can be entirely characterized by the time steps at which each node gets infected. 
We perform those agent-based simulations using $15\%$ of the number of nodes chosen as seed nodes and 100 simulations for each seed node selected, those conditions are repeated for each sparsification considered.
In order to quantify and compare how the spreading process occurs in the original network and in the sparsified networks, we consider two measures that assess the micro and macro dynamics of each node and the larger population, respectively.

In order to evaluate how similarly the infection spreads across the networks at the scale of each node we compare the average order at which the nodes get infected during spreading, $\rho^t$.
That is, the Spearman rank correlation between the time in which nodes get infected in the original graph, $t_i^G$, and in a given sparsified subgraph, $t_i$.
For a given seed node, $\rho^t$ is computed considering the average infection time over 100 simulations and for each amount of sparsification, $\mathcal{X}$. 
We report the average and standard deviation over the seed nodes.

In order to estimate the timescale of the infection outbreak, the macro dynamics, we consider the time at which half of the nodes are infected, $t_{0.5}$, also known as half-infection time \cite{starnini2013immunization}. 
Although analogous quantities referring to different population proportions could be considered, the half-infection time was shown to be less prone to large fluctuations when compared to the full-infection time ($t_{1}$), for example \cite{correia2023contact}.
With this, we aimed at understanding whether the different sparsification methods removed edges that were critical for the global timescale of the infection propagation or not.
In order to appropriately assess this question we considered the ratio between the half-infection time in the original network, $t^{G}_{0.5}$, and in a given sparsified subgraph, $t_{0.5}$, as
\begin{equation}
\xi^{-1}_{0.5} = \frac{t^{G}_{0.5}}{t_{0.5}},
\end{equation}
the inverse of $\xi_{0.5}$ defined in \cite{soriano2025quantifying}.
This way, we have a bounded and normalized measure, $\xi^{-1} \in [0, 1]$, apart from small random fluctuations, since the spreading tends to take longer to propagate in sparser networks.
For each seed node, the median of $\xi^{-1}_{0.5}$ over 100 simulations is taken and we report the mean and standard deviation over different seed nodes.

\subsubsection{Area Under the Curve (AUC) Summarization}
\label{sec:auc_summary_methods}

In order to quantify which method generally performs better, we compute the area under the curves of $\rho^C$, $\rho^t$, and $\xi^{-1}_{0.5}$ for each network until a sparsification limit $\tilde{\mathcal{X}}$.
\begin{equation}
    \textit{AUC}(\rho^C) = \frac{\int^{1}_{\Tilde{\mathcal{X}}}  \rho^C(\mathcal{X}) \: d\mathcal{X}}{1 - \Tilde{\mathcal{X}}} \:\:\:\:\:\:\:\:\:\:\:\:\:\:\:\:
    \textit{AUC}(\rho^t) = \frac{\int^{1}_{\tilde{\mathcal{X}}}  \rho^t(\mathcal{X}) \: d\mathcal{X}}{1 - \tilde{\mathcal{X}}} \:\:\:\:\:\:\:\:\:\:\:\:\:\:\:\: 
    \textit{AUC}(\xi^{-1}_{0.5}) = \frac{\int^{1}_{\Tilde{\mathcal{X}}} \xi^{-1}_{0.5} (\mathcal{X}) \: d\mathcal{X}}{1 - \Tilde{\mathcal{X}}}
\label{eq:area_under_curve}
\end{equation}

With the distribution of this aggregated measure we can evaluate how well a given sparsification method preserves a given performance metric over a reasonable sparsification range in different networks. 
The $\textit{AUC}$ will be closer to $1$ whenever the given sparsification method allows a greater sparsification range without breaking the network (see \ref{sec:sparsification_schemes}) and does it maintaining a good performance metric (large values of either $\rho^C$, $\rho^t$, and $\xi^{-1}_{0.5}$).
Generally, we consider $\tilde{\mathcal{X}}$ to be the smallest sparsification which preserves connectivity; however, when this limit is very small (as is the case for $\textbf{DBS}$), we fix $\tilde{\mathcal{X}}=|B^{m}|$ as a more reasonable sparsification limit size.

\subsection{Optimization of Performance Metrics}
\label{sec:optimization_methods}

\if
In order to identify the optimal backbone under different scenarios, we find the value of $\lambda^D$ which maximizes the objective function
\begin{equation}
    p_C \rho^C + p_t \rho^t + p_{\xi}\xi^{-1}_{0.5} + p_{\mathcal{X}}(1-\mathcal{X})
\end{equation}
constraint to $p_C + p_t + p_{\xi} + p_{\mathcal{X}} = 1$ for different values of $p_C,\, p_t,\, p_{\xi} \in (0, 1)$ according to the scenarios in Table~\ref{tab:optimization_scenarios}.

\begin{table}[h!]
    \centering
    \caption{Objective functions coefficients for the different optimization scenarios.}
    \begin{tabular}{|c||c|c|c|c|}
        \hline
        Scenario Name & $p_C$ & $p_t$ & $p_{\xi}$ & $p_{\mathcal{X}}$ \\
        \hline\hline
        Balanced & 0.25 & 0.25 & 0.25 & 0.25\\ \hline
        Balanced Except $\xi^{-1}_{0.5}$ & 1/3 & 1/3 & 0 & 1/3 \\ \hline
        Balanced Except $\rho^t$ & 1/3 & 0 & 1/3 & 1/3\\ \hline
        Balanced Except $\rho^C$ & 0 & 1/3 & 1/3 & 1/3 \\ \hline
        Balanced Except $\mathcal{X}$ & 1/3 & 1/3 & 1/3 & 0 \\ \hline
        Slight Preference $\xi^{-1}_{0.5}$ & 0.2 & 0.2 & 0.4 & 0.2 \\ \hline
        Slight Preference $\rho^t$ & 0.2 & 0.4 & 0.2 & 0.2 \\ \hline
        Slight Preference $\rho^C$ & 0.4 & 0.2 & 0.2 & 0.2 \\ \hline
        Slight Preference $\mathcal{X}$ & 0.2 & 0.2 & 0.2 & 0.4 \\ \hline
    \end{tabular}
    ~
    \begin{tabular}{|c||c|c|c|c|}
        \hline
        Scenario Name & $p_C$ & $p_t$ & $p_{\xi}$ & $p_{\mathcal{X}}$ \\
        \hline\hline
        Strong Preference $\xi^{-1}_{0.5}$ & 0.1 & 0.1 & 0.7 & 0.1 \\ \hline
        Strong Preference $\rho^t$ & 0.1 & 0.7 & 0.1 & 0.1 \\ \hline
        Strong Preference $\rho^C$ & 0.7 & 0.1 & 0.1 & 0.1 \\ \hline
        Strong Preference $\mathcal{X}$ & 0.1 & 0.1 & 0.1 & 0.7\\ \hline
        Mainly $\xi^{-1}_{0.5}$ & 0.02 & 0.02 & 0.94 & 0.02 \\ \hline
        Mainly $\rho^t$ & 0.02 & 0.94 & 0.02 & 0.02 \\ \hline
        Mainly $\rho^C$ & 0.94 & 0.02 & 0.02 & 0.02 \\ \hline
        Mainly $\mathcal{X}$ & 0.02 & 0.02 & 0.02 & 0.94 \\ \hline
        \multicolumn{5}{c}{}
    \end{tabular}
    \label{tab:optimization_scenarios}
\end{table}
\fi

In order to identify the optimal backbone under different scenarios, we find the value of $\lambda^D$ which maximizes the objective function
\begin{equation}
    w_C \rho^C + w_t \rho^t + w_{\xi}\xi^{-1}_{0.5} + w_{\mathcal{X}}(1-\mathcal{X})
\end{equation}
constraint to $w_C + w_t + w_{\xi} + w_{\mathcal{X}} = 1$ for different values of $w_C,\, w_t,\, w_{\xi} \in (0, 1)$ according to the scenarios in Table~\ref{tab:optimization_scenarios}.

\begin{table}[h!]
    \centering
    \caption{Objective functions coefficients for the different optimization scenarios.}
    \begin{tabular}{|c||c|c|c|c|}
        \hline
        Scenario Name & $w_C$ & $w_t$ & $w_{\xi}$ & $w_{\mathcal{X}}$ \\
        \hline\hline
        Balanced & 0.25 & 0.25 & 0.25 & 0.25\\ \hline
        Balanced Except $\xi^{-1}_{0.5}$ & 1/3 & 1/3 & 0 & 1/3 \\ \hline
        Balanced Except $\rho^t$ & 1/3 & 0 & 1/3 & 1/3\\ \hline
        Balanced Except $\rho^C$ & 0 & 1/3 & 1/3 & 1/3 \\ \hline
        Balanced Except $\mathcal{X}$ & 1/3 & 1/3 & 1/3 & 0 \\ \hline
        Slight Preference $\xi^{-1}_{0.5}$ & 0.2 & 0.2 & 0.4 & 0.2 \\ \hline
        Slight Preference $\rho^t$ & 0.2 & 0.4 & 0.2 & 0.2 \\ \hline
        Slight Preference $\rho^C$ & 0.4 & 0.2 & 0.2 & 0.2 \\ \hline
        Slight Preference $\mathcal{X}$ & 0.2 & 0.2 & 0.2 & 0.4 \\ \hline
    \end{tabular}
    ~
    \begin{tabular}{|c||c|c|c|c|}
        \hline
        Scenario Name & $w_C$ & $w_t$ & $w_{\xi}$ & $w_{\mathcal{X}}$ \\
        \hline\hline
        Strong Preference $\xi^{-1}_{0.5}$ & 0.1 & 0.1 & 0.7 & 0.1 \\ \hline
        Strong Preference $\rho^t$ & 0.1 & 0.7 & 0.1 & 0.1 \\ \hline
        Strong Preference $\rho^C$ & 0.7 & 0.1 & 0.1 & 0.1 \\ \hline
        Strong Preference $\mathcal{X}$ & 0.1 & 0.1 & 0.1 & 0.7\\ \hline
        Mainly $\xi^{-1}_{0.5}$ & 0.02 & 0.02 & 0.94 & 0.02 \\ \hline
        Mainly $\rho^t$ & 0.02 & 0.94 & 0.02 & 0.02 \\ \hline
        Mainly $\rho^C$ & 0.94 & 0.02 & 0.02 & 0.02 \\ \hline
        Mainly $\mathcal{X}$ & 0.02 & 0.02 & 0.02 & 0.94 \\ \hline
        \multicolumn{5}{c}{}
    \end{tabular}
    \label{tab:optimization_scenarios}
\end{table}

\section{Acknowledgements}

We thank the members of the CASCI lab for useful discussions about this project. 
This work is financed by Portuguese national funds through the FCT - Foundation for Science and Technology, I.P., under the project 2022.09122.PTDC (\href{https://doi.org/10.54499/2022.09122.PTDC}{https://doi.org/10.54499/2022.09122.PTDC}).
LMR was also partially funded by NIH National Library of Medicine Program grant 01LM011945-01 and the European Union’s Horizon Europe research and innovation programme under Grant Agreement No. 101186558 for the ‘Strategic Integration of Complex Networks and Systems for Advancing Biomedical Research’.
The funders had no role in any stage of the study design to the manuscript preparation, nor decision to publish.

\bibliographystyle{unsrt}  
\bibliography{references} 

\end{document}


\appendix
\section{Implementation of the Distance Backbone Synthesis}
\label{sec:implementation}

Algorithms to identify a distance backbone can be implemented in several ways \cite{simas2021distance, kalavri2016shortest}.
One option is to first construct the corresponding distance closure and then compare the edge weights with the original distance graph.
This construction will depend on the quantification of path lengths according to $g$, a \emph{triangular distance norm} (td-norm) \cite{simas2021distance, simas2015distance}.
As introduced in Sec.\ref{main:sec:background}, the adjacency matrix elements of the distance closure graph due to the td-norm $g$ will be
\begin{equation}
    d_{ij}^{T, g} = \underset{\Gamma}{\min} \: g(d_{ik_1}, \dots, d_{k_nj}), 
\end{equation}
where $\Gamma = (x_i, x_{k_1}, \dots, x_{k_n}, x_j)$ is a path along the graph, $G$, with distance adjacency matrix elements $d_{ij}$.
Then, the edges in the distance backbone $B^g$ will be those satisfying $d^{T,g}_{ij} = d_{ij}$.

Another option to identify the edges in $B^g$ is in terms of an isomorphic proximity adjacency matrix elements, $p_{ij}$.
After all, td-norms originate from an isomorphic transformation of \emph{triangular norms} (t-norms) \cite{simas2015distance}, which themselves have been originally described in the context of probabilistic metric spaces \cite{menger1942statistical} and fuzzy logic \cite{novak1999mathematical}.
That is, a t-norm $\wedge$ can be found by the operational composition
\begin{equation}
    \wedge = \varphi\inv \circ g \circ \varphi,
\end{equation}
where $\varphi:[0, 1] \to [0, \infty]$ maps the edge weights in terms of $p_{ij}$ to $d_{ij}$.
Interestingly, relations in fuzzy logic are commonly represented as directed proximity graphs known as fuzzy graphs \cite{klir1995fuzzy} and t-norms establish a transitivity criteria for them.
This criteria allow us to compute the transitive closure of $G$ which will have proximity adjacency matrix elements 
\begin{equation}
    p_{ij}^{T, \wedge} = \underset{\Gamma}{\max} \: \wedge(p_{ik_1}, \dots, p_{k_nj}), 
    \label{eq:pclosure}
\end{equation}
and the edges in the backbone $B^g$ will also satisfy $p^{T,\wedge}_{ij} = p_{ij}$.
The $\min$ operator for distance measures turns into a $\max$ operator for proximity measures because $\varphi$ is a continuous strictly decreasing map with $\varphi(0)=+\infty$ and $\varphi(1)=0$.
Furthermore, as $\wedge$ establishes a transitivity criteria for graphs $G$ in terms of proximity, $g$ is also responsible for the transitivity of distance closures, explaining the superscript notation $T$.

Lastly, there is a more efficient way to find a distance backbones, $B^g$, which is implemented in our \textsc{distanceclosure} package \cite{distanceclosure2025} and used in this work.
Our algorithm solves single source shortest path problems via Dijkstra \cite{brandes2005network}, accounting for the path length measure $g$, with iterative edge removal per source node.
An edge is removed if $d^{T,g}_{ij} < d_{ij}$, which ultimately results in keeping only those with $d^{T,g}_{ij} = d_{ij}$.
This way, the distance backbone, $B^g$, is iteratively found without the need for an extra comparison with the overall distance closure.

\begin{figure}[t!]
\centering
\begin{tikzpicture}[scale=1.5]

\node (P) at (0, 0) {$p_{ij}$};
\node (D) at (2, 0) {${d'}_{ij}$};
\node (PT) at (0, -2) {$p^{T,\wedge_\lambda}_{ij}$};
\node (DT) at (2, -2) {${d'}_{ij}^{T, +}$};
\node (D2) at (-2, 0) {$d_{ij}$};
\node (DT2) at (-2, -2) {$d_{ij}^{T, g_{\lambda}}$};

\draw[<-, color = red] ([yshift=0.5mm] D2.east) -- ([yshift=0.5mm] P.west) node[midway, above] {$\varphi$};
\draw[->] ([yshift=-0.5mm] D2.east) -- ([yshift=-0.5mm] P.west) node[midway, below] {$\varphi\inv$};
\draw[<-] ([yshift=0.5mm] DT2.east) -- ([yshift=0.5mm] PT.west) node[midway, above] {$\varphi$};
\draw[->, , color = red] ([yshift=-0.5mm] DT2.east) -- ([yshift=-0.5mm] PT.west) node[midway, below] {$\varphi\inv$};
\draw[->, color = red] (D2) -- (DT2) node[midway, left] {$\min \circ \:g_{\lambda}$};
\draw[->, color = blue] ([yshift=0.5mm] P.east) -- ([yshift=0.5mm] D.west) node[midway, above] {$\varphi_\lambda$};
\draw[<-] ([yshift=-0.5mm] P.east) -- ([yshift=-0.5mm] D.west) node[midway, below] {$\varphi_\lambda\inv$};
\draw[->] ([yshift=0.5mm] PT.east) -- ([yshift=0.5mm] DT.west) node[midway, above] {$\varphi_\lambda$};
\draw[<-, color = blue] ([yshift=-0.5mm] PT.east) -- ([yshift=-0.5mm] DT.west) node[midway, below] {$\varphi_\lambda\inv$};
\draw[->] (P) -- (PT) node[midway, left] {$\max \circ \: \wedge_\lambda$};
\draw[->, color = blue] (D) -- (DT) node[midway, right] {$\min \circ \: +$};

\end{tikzpicture}
\caption{Isomorphisms between closures from a family of triangular norms ($\wedge_\lambda$) to a family of triangular distance norms ($g_\lambda$) or maps ($\varphi_\lambda$).}
\label{fig:prox_dist_isomorphism}
\end{figure}
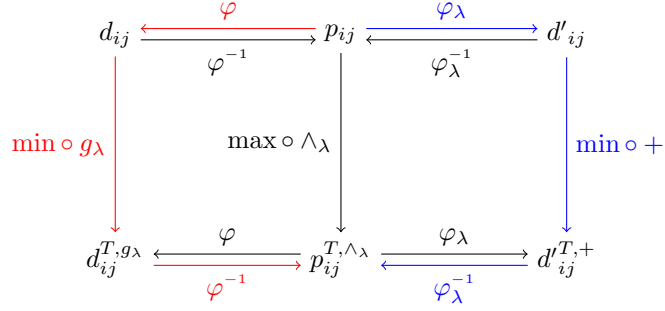

Choosing an appropriate isomorphic transformation, $\varphi$, can facilitate the implementation of the distance backbone synthesis.
Since our original networks are naturally represented by graphs with proximity edge weights $p_{ij}\in[0, 1]$, the ``natural'' way to identify a backbone $B^{g_\lambda}$ induced by the td-norm $g_\lambda$ (See Sec.~\ref{main:sec:dbs_intro}) would be in terms of the t-norm
\begin{equation}
    \wedge_\lambda = \varphi\inv \circ g_\lambda \circ \varphi = \frac{xy}{xy + \qty[ y^\lambda(1-x)^\lambda + x^\lambda(1-y)^\lambda ]^{\frac{1}{\lambda}}},
    \label{eq:DombiTnorm}
\end{equation}
where we used the isomorphism $\varphi = \frac1x - 1$ from Eq.~\ref{main:eq:isomorphism}, which was introduced in previous studies \cite{simas2021distance, costa2023distance}. 
In other words, $B^{g_\lambda}$ contains the edges satisfying $p_{ij} = p^{T,\wedge_\lambda}_{ij}$ as defined in Equations~\ref{eq:pclosure} and~\ref{eq:DombiTnorm}.
However, as previously mentioned, it is more efficient to iteratively find $B^{g_\lambda}$ using the Dijkstra algorithm accounting for the path length measure $g_\lambda$.
This can be done without altering the Dijkstra algorithm by taking advantage of an alternative isomorphism
\begin{equation}
    \varphi_\lambda = \qty(\frac1x - 1)^\lambda,
    \label{eq:dombi_isomorphism}
\end{equation}
which leads to the path length measure, $g' = \varphi_\lambda \circ \wedge_\lambda \circ \varphi_\lambda\inv \equiv +$, being the traditional sum.
This parametrized isomorphism, is known in the fuzzy logic literature as \textit{additive decreasing generators} \cite{klir1995fuzzy, klement2000triangular}.
In summary, identifying which edges are in $B^{g_\lambda}$ can be done using the canonical mapping $\varphi$ \cite{simas2021distance} and path-length measure $g_\lambda$ (left side of Fig.\ref{fig:prox_dist_isomorphism}); or equivalently the specific mapping $\varphi_\lambda$ in which the path-length measure is a sum (right side of Fig.\ref{fig:prox_dist_isomorphism}).

Since both methodologies produce distance closures isomorphic to each-other, by being isomorphic to an intermediate proximity closure, we conclude that the underlying distance backbones are also isomorphic, and thus contain exactly the same edges.
Nevertheless, this feature is mostly relevant when the original graphs contain similarity data (edge weights as $p_{ij} \in [0,1]$). Otherwise, in the cases where the graph originally is built from dissimilarity data (edge weights as $d_{ij} \in [0,+\infty]$), directly applying $g_{\lambda}$ is more appropriate to apply the distance backbone synthesis (\textbf{DBS}).
For practical reasons, and since both produce the same backbones, in the experimental part of this work we used the right-hand side of the diagram in Fig. \ref{fig:prox_dist_isomorphism}.
However, we recurred to the explanation of the left-hand side in the main text as it is more clearly understandable and sufficient for other scenarios such as when the weights are already distances.

In order to compute the parameter below which an edge is part of a backbone, $\lambda_{ij}$, we make use of the following two algorithms.
The first one (Algo.~\ref{alg:is_edge_in_backbone}) checks if a given edge $(x_i,x_j)$ is in the distance backbone $B^{g_\lambda}$ associated with a given $\varphi_{\lambda}$, i.e. if it is a shortest-path in the resulting distance graph. 
This algorithm requires to isomorphically transform the edge weights from proximity to distance because it receives as input the original proximity graph.

\begin{algorithm}[H]
\textbf{Input:} $(x_i,x_j)$ (edge), $\{p_{ij}\}$ (graph with proximity edge weights), $\varphi_{\lambda}$ (specific proximity-distance isomorphism)\\
\\
\phantom{0}1. $d_{ij} \gets \varphi_{\lambda}(p_{ij})$\\
\phantom{0}2. $\textit{sp} \gets \texttt{shortest-path}(\{d_{ij}\}, (x_i, x_j), +)$\\
\phantom{0}3. \textbf{IF} {$\textit{sp}=(x_i,x_j)$}:\\
\phantom{0}4. \hspace{5mm} \textbf{RETURN} \texttt{True}\\
\phantom{0}5. \textbf{ELSE}:\\
\phantom{0}6. \hspace{5mm} \textbf{RETURN} \texttt{False}
\caption{\texttt{is\_edge\_in\_backbone}}
\label{alg:is_edge_in_backbone}
\end{algorithm}

The second one (Algo.~\ref{alg:compute_lambdas}) searches the space of $\lambda \in [0,+\infty]$ for each edge $(x_i,x_j)$ to find the corresponding $\lambda_{ij}$. 
This is done using binary search and setting a lower and an upper bound in that space, $(\lambda_{\min}, \lambda_{\max})$, and the number of times we want to divide that interval in the search for $\lambda_{ij}$, $N$.
\begin{algorithm}
\textbf{Input:} $\{p_{ij}\}$ (graph with proximity weights), $\varphi_{\lambda}$ (family of proximity-distance isomorphisms)\\
\phantom{Input:} $\lambda_{\min}$ (Lower Bound), $\lambda_{\max}$ (Upper Bound), $N$ (number of iterations)\\
\\
\phantom{0}1. $\Lambda \gets \{ (x_i,x_j): \texttt{None} \mid  p_{ij} \in (0, 1)\}$\\
\phantom{0}2. \textbf{FOR EACH} $(i,j) \in \{p_{ij}\}$:\\
\phantom{0}3. \hspace{5mm} $L \gets \texttt{is\_edge\_in\_backbone}((x_i,x_j), \{p_{ij}\}, \varphi_{\lambda_{\min}})$\\
\phantom{0}4. \hspace{5mm} $R \gets \texttt{is\_edge\_in\_backbone}((x_i,x_j), \{p_{ij}\}, \varphi_{\lambda_{\max}})$\\
\phantom{0}5. \hspace{5mm} \textbf{IF} $L=\texttt{True}$ \textbf{AND} $R=\texttt{False}$ :\\
\phantom{0}6. \hspace{10mm} \textbf{FOR EACH} $n \in \textbf{RANGE}(N)$:\\
\phantom{0}7. \hspace{15mm} $\lambda_C \gets (\lambda_{\min}+\lambda_{\max})/2$\\
\phantom{0}8. \hspace{15mm} $C \gets \texttt{is\_edge\_in\_backbone}((x_i,x_j), \{p_{ij}\}, \varphi_{\lambda_{C}})$\\
\phantom{0}9. \hspace{15mm} \textbf{IF} $C=\texttt{True}$: $\lambda_{\min} \gets \lambda_C$\\
10. \hspace{15mm} \textbf{ELSE} $\lambda_{\max} \gets \lambda_C$\\
11. \hspace{10mm} $\Lambda[(x_i, x_j)]\gets \lambda_{\min}$ \\
12. \hspace{5mm} \textbf{ELSE}:\\
13. \hspace{10mm} $\Lambda[(x_i,x_j)] \gets \texttt{NOT FOUND}$\\
14. \textbf{RETURN} $\Lambda$
\caption{\texttt{compute\_lambdas}}
\label{alg:compute_lambdas}
\end{algorithm}

In this work we set $\lambda_{\min}=10^{-4}$, $\lambda_{\max}=10^3$ and $N=50$.
Therefore, the value found for each edge is associated with a precision given by:
\begin{equation}
    \epsilon = \frac{\lambda_{\max} - \lambda_{\min}}{2^N} = \frac{10^3 - 10^{-4}}{2^{50}} \approx 8.88\times 10^{-13}
\end{equation}
Thus, with such an infinitesimal precision, throughout this work we take every value of $\lambda_{ij}$ as exact and not approximate.

In exceptional cases, the bounds provided might not yield a $\lambda_{ij}$.
This can happen if the edge is in the backbone for both values, for none, or even due to floating point operation errors. 
In these cases a more customized search is necessary through the change of the initial bounds, which only happened for 2 of the networks we considered. 
Additionally, the \texttt{shortest-path} computation can be done using any commonly known algorithm, such as Dijkstra.
In this case, the overall computational complexity of this algorithm is $\mathcal{O}(|E| + |E| \cdot N \cdot \text{Dijkstra})$.

However, as mentioned before, the same values of $\lambda_{ij}$ could be obtained by using the algorithms performing the computations of the left-hand side of Fig.\ref{fig:prox_dist_isomorphism}. 
In this scenario, the \texttt{shortest-path} computation must use $g_{\lambda}$ as the path-length measure and the above mentioned algorithms become Algo.\ref{alg:is_edge_in_backbone2} and Algo.
\ref{alg:compute_lambdas2}.

\begin{algorithm}[H]
\textbf{Input:} $(x_i,x_j)$ (edge), $\{d_{ij}\}$ (graph with distance weights), $g_{\lambda}$ (specific \textit{td-norm})\\
\\
\phantom{0}1. $\textit{sp} \gets \texttt{shortest-path}(\{d_{ij}\}, (x_i,x_j), g_{\lambda})$\\
\phantom{0}2. \textbf{IF} {$\textit{sp}=(x_i,x_j)$}:\\
\phantom{0}3. \hspace{5mm} \textbf{RETURN} \texttt{True}\\
\phantom{0}4. \textbf{ELSE}:\\
\phantom{0}5. \hspace{5mm} \textbf{RETURN} \texttt{False}
\caption{\texttt{is\_edge\_in\_backbone}}
\label{alg:is_edge_in_backbone2}
\end{algorithm}

\begin{algorithm}[H]
\textbf{Input:} $\{d_{ij}\}$ (graph with distance weights), $g_{\lambda}$ (family of \textit{td-norms})\\
\phantom{Input:} $\lambda_{\min}$ (Lower Bound), $\lambda_{\max}$ (Upper Bound), $N$ (number of iterations)\\
\\
\phantom{0}1. $\Lambda \gets \{ (x_i,x_j): \texttt{None} \mid d_{i,j} \in (0, \infty)\}$\\
\phantom{0}2. \textbf{FOR EACH} $(x_i,x_j) \in \{d_{ij}\}$:\\
\phantom{0}3. \hspace{5mm} $L \gets \texttt{is\_edge\_in\_backbone}((x_i,x_j), \{d_{ij}\}, g_{\lambda_{\min}})$\\
\phantom{0}4. \hspace{5mm} $R \gets \texttt{is\_edge\_in\_backbone}((x_i,x_j), \{d_{ij}\}, g_{\lambda_{\max}})$\\
\phantom{0}5. \hspace{5mm} \textbf{IF} $L=\texttt{True}$ \textbf{AND} $R=\texttt{False}$ :\\
\phantom{0}6. \hspace{10mm} \textbf{FOR EACH} $n \in \textbf{RANGE}(N)$:\\
\phantom{0}7. \hspace{15mm} $\lambda_C \gets (\lambda_{\min}+\lambda_{\max})/2$\\
\phantom{0}8. \hspace{15mm} $C \gets \texttt{is\_edge\_in\_backbone}((x_i,x_j), \{d_{ij}\}, g_{\lambda_{C}})$\\
\phantom{0}9. \hspace{15mm} \textbf{IF} $C=\texttt{True}$: $\lambda_{\min} \gets \lambda_C$\\
10. \hspace{15mm} \textbf{ELSE} $\lambda_{\max} \gets \lambda_C$\\
11. \hspace{10mm} $\Lambda[(x_i,x_j)]\gets \lambda_{\min}$ \\
12. \hspace{5mm} \textbf{ELSE}:\\
13. \hspace{10mm} $\Lambda[(x_i,x_j)] \gets \texttt{NOT FOUND}$\\
14. \textbf{RETURN} $\Lambda$
\caption{\texttt{compute\_lambdas}}
\label{alg:compute_lambdas2}
\end{algorithm}

\section{Distance Backbone Synthesis from other \textit{td-norm} families}
\label{sec:other_synthesis}

\begin{table}[h!]
    \centering
    \caption{Characterization of studied families capable of synthesizing graphs by their distance backbones.}
    \begin{tabular}{c|c|c|c}
         & \textbf{Path-length measure}, $g_\lambda$ & \textbf{T-norm}, $\wedge_\lambda$ & \textbf{Isomorphism}, $\varphi_\lambda$ \\
        \hline
        \begin{sideways} \textbf{Dombi}\, \end{sideways} & $\qty( x^\lambda + y^\lambda)^{\frac1\lambda}$ & $\dfrac{xy}{xy + \qty[ y^\lambda(1-x)^\lambda + x^\lambda(1-y)^\lambda ]^{\frac{1}{\lambda}}}$ & $\qty(\frac1x - 1)^\lambda$ \\ \hline
        \begin{sideways} \textbf{Aczél-Alsina}\, \end{sideways} & $\exp{\log^\lambda\qty(1+x) + \log^\lambda\qty(1+y)}-1$  & $\exp{-\qty[\log^\lambda\qty(\frac1x)+\log^\lambda\qty(\frac1y)]^{\frac1\lambda}}$ & $\log^\lambda\qty(\frac1x)$ \\ \hline
    \end{tabular}
    \label{tab:norm_families}
\end{table}

The Distance Backbone Synthesis ($\textbf{DBS}$) allows to sweep different topological spaces defined by a path length measure $g_\lambda$, also known as \textit{td-norm}, from $g_{\text{max}}$ to $g_{\text{drastic}}$.
In the main text, we focused on the Dombi family for it includes the traditional sum operation for distances (Eq.~\ref{main:eq:dombi_tdnorms}).
This family inherit its name because applying the traditional isomorphism (Eq.~\ref{main:eq:isomorphism}) to it, one finds the Dombi triangular norm \cite{dombi1982general, klir1995fuzzy}.
Furthermore, the Dombi \textit{additive decreasing generators} is the proximity-distance family of isomorphisms $\varphi_\lambda$ defined in Eq.~\ref{eq:dombi_isomorphism}.
That is, the characterization of families capable of synthesizing the topological spaces of edges via the shortest path backbones is done by defining a path length measure $g_\lambda$, or equivalently a triangular norm $\wedge_\lambda$ with an accompanying \textit{additive decreasing generators} $\varphi_\lambda$.
Those operations are shown in Table~\ref{tab:norm_families} for the families studied here and are also available in the literature \cite{klir1995fuzzy, klement2000triangular}.
Other families have also been previously characterized \cite{pereira2023distance}.

\begin{figure}[h]
    \centering
    \includegraphics[width=\linewidth]{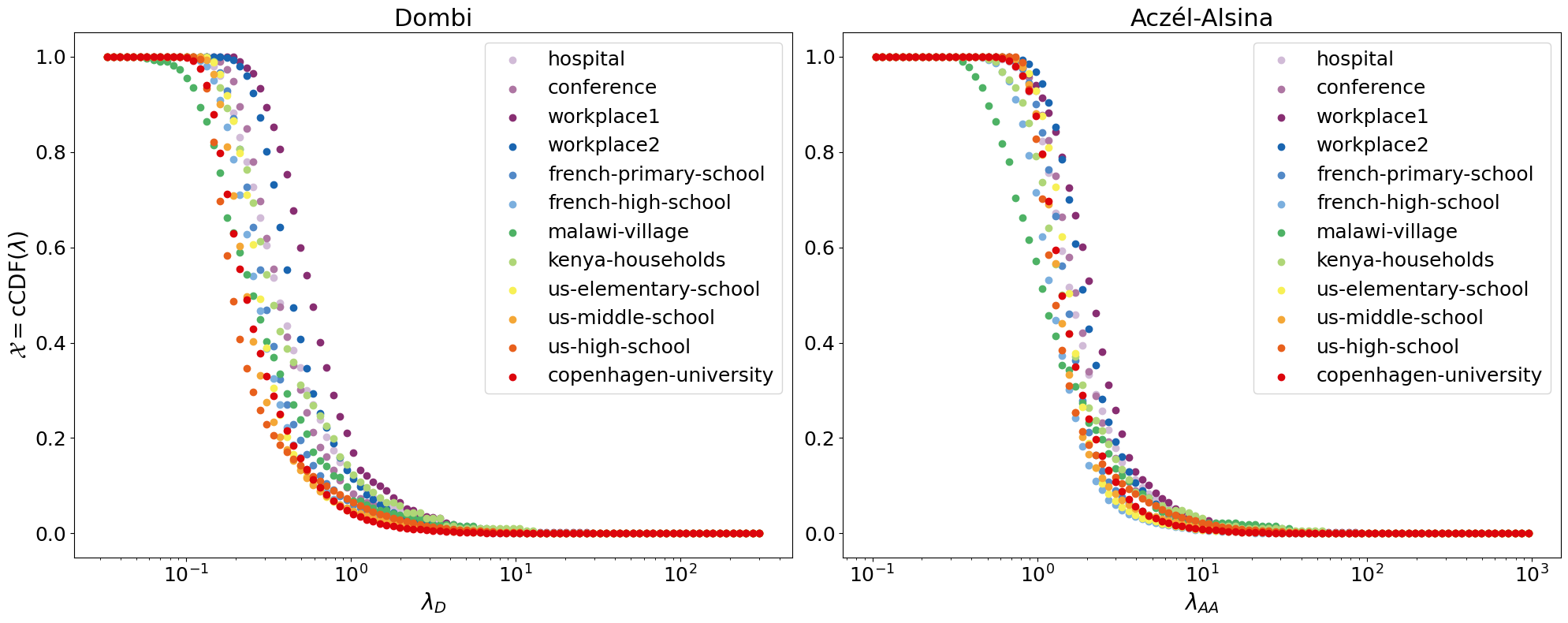}
    \caption{Adjusted subgraph size for DBS under different families.}
    \label{fig:families_chi_lambda}
\end{figure}

It is interesting to note that for both Dombi and Aczél-Alsina families of distance backbones, the complementary cumulative distribution function of $\lambda_{ij}$ has a sigmoidal shape in semi-log scale, suggesting a log-normally distributed $\lambda_{ij}$, as shown in Fig.~\ref{fig:families_chi_lambda}.
This is further corroborated by a log-normal curve-fit of $\mathcal{X}(\lambda)$ for both families which has a perfect $R^2$ for all cases considered (Tables~\ref{tab:dombi_lambda_dist} and~\ref{tab:aczel_alsina_lambda_dist}).
Furthermore, comparing both families show that the natural topology of most edges is between the product backbone, found by setting $\lambda^{AA} = 1$, and the metric backbone, found by setting $\lambda^D = 1$.
This is observed in the mode of the $\lambda^D$ distribution being $0.31\pm0.01$ (Tables~\ref{tab:dombi_lambda_dist}) and the mode of the $\lambda^{AA}$ distribution being $1.46\pm0.27$.

\begin{table}[h!]
    \centering
    \caption{Description for $\lambda^D$ distribution fit.}

    \begin{tabular}{l||rr||rrrr}
    \toprule
    Network & $\min(\lambda^D)$ & $\max(\lambda^D)$ & $R^2$ (cCDF) & $\sigma$ & $\mu$ & mode \\
    \midrule
    hospital & 0.133 & 16.276 & 0.996 & 0.387 & -0.937 & 0.337 \\
    conference & 0.142 & 18.365 & 0.996 & 0.308 & -0.961 & 0.348 \\
    workplace1 & 0.207 & 39.003 & 0.998 & 0.323 & -0.524 & 0.533 \\
    workplace2 & 0.159 & 24.593 & 0.996 & 0.296 & -0.764 & 0.427 \\
    french-primary-school & 0.122 & 31.919 & 0.996 & 0.277 & -1.157 & 0.291 \\
    french-high-school & 0.097 & 13.436 & 0.996 & 0.281 & -1.265 & 0.261 \\
    malawi-village & 0.058 & 13.563 & 0.995 & 0.439 & -1.293 & 0.227 \\
    kenya-households & 0.146 & 14.182 & 0.994 & 0.405 & -0.982 & 0.318 \\
    us-elementary-school & 0.122 & 27.489 & 0.996 & 0.232 & -1.236 & 0.275 \\
    us-middle-school & 0.119 & 162.773 & 0.993 & 0.248 & -1.389 & 0.234 \\
    us-high-school & 0.115 & 160.516 & 0.985 & 0.279 & -1.543 & 0.198 \\
    copenhagen-university & 0.088 & 128.818 & 0.995 & 0.319 & -1.384 & 0.226 \\
    \bottomrule
    mean & 0.126 & 54.244 & 0.995 & 0.316 & -1.119 & 0.306 \\
    std & 0.038 & 59.248 & 0.003 & 0.063 & 0.293 & 0.096 \\
    \end{tabular}    
    \label{tab:dombi_lambda_dist}
\end{table}
~
\begin{table}[h!]
    \centering
    \caption{Description for $\lambda^{AA}$ distribution fit.}

    \begin{tabular}{l||rr||rrrr}
    \toprule
    Network & $\min(\lambda^{AA})$ & $\max(\lambda^{AA})$ & $R^2$ (cCDF) & $\sigma$ & $\mu$ & mode \\
    \midrule
    hospital & 0.594 & 60.474 & 0.997 & 0.316 & 0.536 & 1.547 \\
    conference & 0.547 & 59.174 & 0.998 & 0.272 & 0.572 & 1.644 \\
    workplace1 & 0.736 & 123.715 & 0.999 & 0.304 & 0.769 & 1.967 \\
    workplace2 & 0.633 & 78.935 & 0.998 & 0.258 & 0.683 & 1.852 \\
    french-primary-school & 0.562 & 103.598 & 0.998 & 0.220 & 0.427 & 1.461 \\
    french-high-school & 0.389 & 41.937 & 0.999 & 0.249 & 0.227 & 1.179 \\
    malawi-village & 0.340 & 43.632 & 0.997 & 0.435 & 0.170 & 0.981 \\
    kenya-households & 0.502 & 56.659 & 0.997 & 0.324 & 0.404 & 1.348 \\
    us-elementary-school & 0.565 & 81.388 & 0.998 & 0.189 & 0.448 & 1.510 \\
    us-middle-school & 0.619 & 350.155 & 0.997 & 0.200 & 0.337 & 1.346 \\
    us-high-school & 0.541 & 460.845 & 0.994 & 0.230 & 0.304 & 1.285 \\
    copenhagen-university & 0.422 & 478.126 & 0.998 & 0.252 & 0.405 & 1.407 \\
    \bottomrule
    mean & 0.537 & 161.553 & 0.997 & 0.271 & 0.440 & 1.461 \\
    std & 0.111 & 166.056 & 0.001 & 0.067 & 0.177 & 0.273 \\
    \end{tabular}    
    \label{tab:aczel_alsina_lambda_dist}
\end{table}

In the main text, we performed the distance backbone synthesis for the Dombi family of path length measures (Eg.~\ref{main:eq:dombi_tdnorms}).
However, we could also evaluate the synthesis performance under the Aczél-Alsina family, by setting $g_\lambda$ as shown in Table~\ref{tab:norm_families}.
In both cases, edges are removed in increasing value of $\lambda_{ij}$ (Tab.~\ref{tab:extra-methods-used}) and in order to distinguish them we refer to \textbf{DBS} as the synthesis via the Dombi family and \textbf{DBS}($\lambda^{\text{AA}}$) via the Aczél-Alsina family.
Another quantifier for edge importance we consider is the semi-ultrametric distortion, $s^{um}_{ij}$, which computes $s^{g}$ (Eq.~\ref{main:eq:distortion_general}) for $g_{\max}$.
Similarly to the semi-metric distortion, a related concept, the sparsification happens in decreasing order of $s^{um}_{ij}$ (Tab.~\ref{tab:extra-methods-used}).
Below we compare the performance of both methodologies and those presented in the main text for all empirical networks studied.

\begin{table}[h!]
    \centering
    \renewcommand{\arraystretch}{1.2}
    \caption{Extra Sparsification Methods Used}
    \begin{tabular}{c|c|c|c}
        \hline
        \textbf{Label} & \textbf{Name} & \textbf{Parameter} & \textbf{Rank Order} \\
        \hline\hline
        \textbf{SumD} & Semi-Ultrametric Distortion & $s^{um}_{ij}$ & increasing ($\nearrow$) \\ \hline
        \textbf{DBS}($\lambda^{\text{AA}}$)  & \textit{Aczél-Alsina Backbones Synthesis}    & $\lambda^{\text{AA}}_{ij}$     & decreasing ($\searrow$) \\ \hline
    \end{tabular}
    \label{tab:extra-methods-used}
\end{table}

\newpage
\section{Additional Results}

\begin{figure}[h!]
\centering
\includegraphics[width=0.8\linewidth]{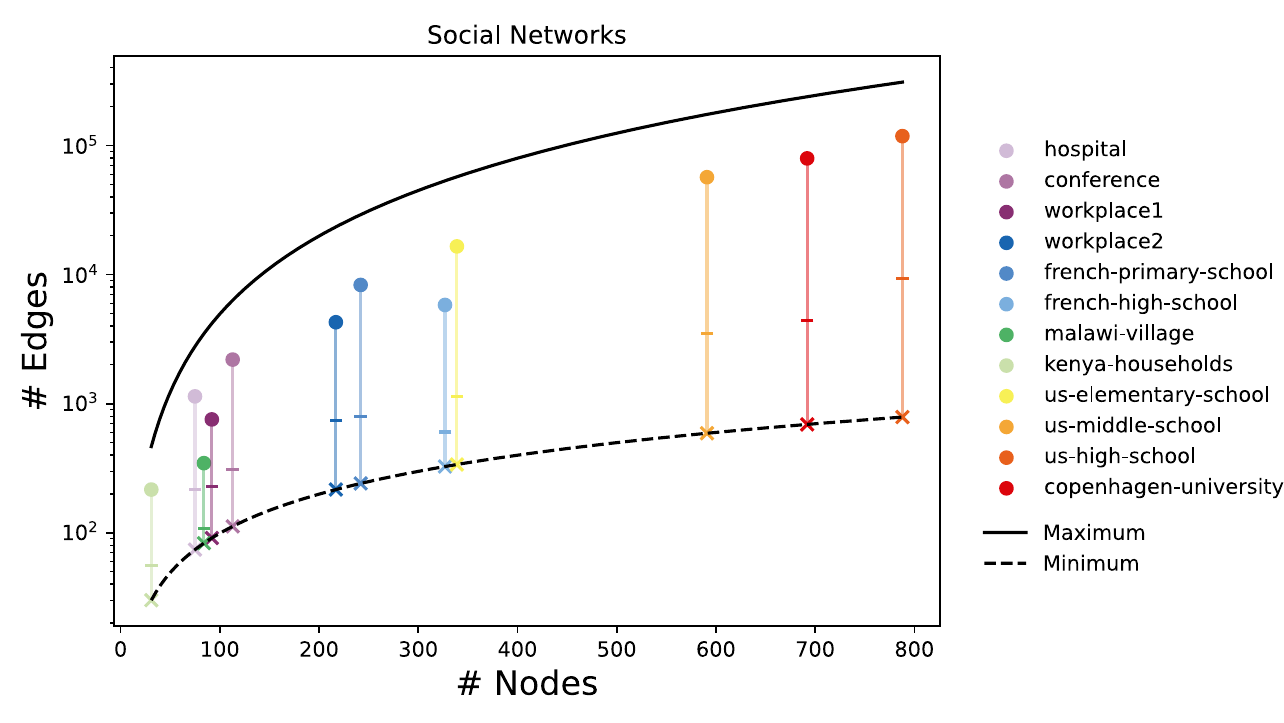}
\caption{Social Networks Density. The black line corresponds to the maximum number of edges in a graph with a certain number of nodes and the dashed line corresponds to the smallest number of edges necessary to have a connected graph. The horizontal dashes in each network line corresponds to the size of the metric backbone and the cross to the ultrametric backbone size.}
\label{fig:network_sizes}
\end{figure}

The analyzed empirical networks span multiple orders of magnitude both in terms of nodes and edges, with less than half of the maximum number of possible connections.
Their sizes can be compared in Fig.~\ref{fig:network_sizes}, which also shows that the ultrametric backbone $B^{um}$ of those networks is its minimum spanning tree, the smallest possible subgraph which doesn't break the network connectivity.

\newpage
\subsection{Preserving dynamically relevant topological measure: Eigenvector Centrality Rank}

\begin{figure}[h!]
\centering
\includegraphics[width=\linewidth]{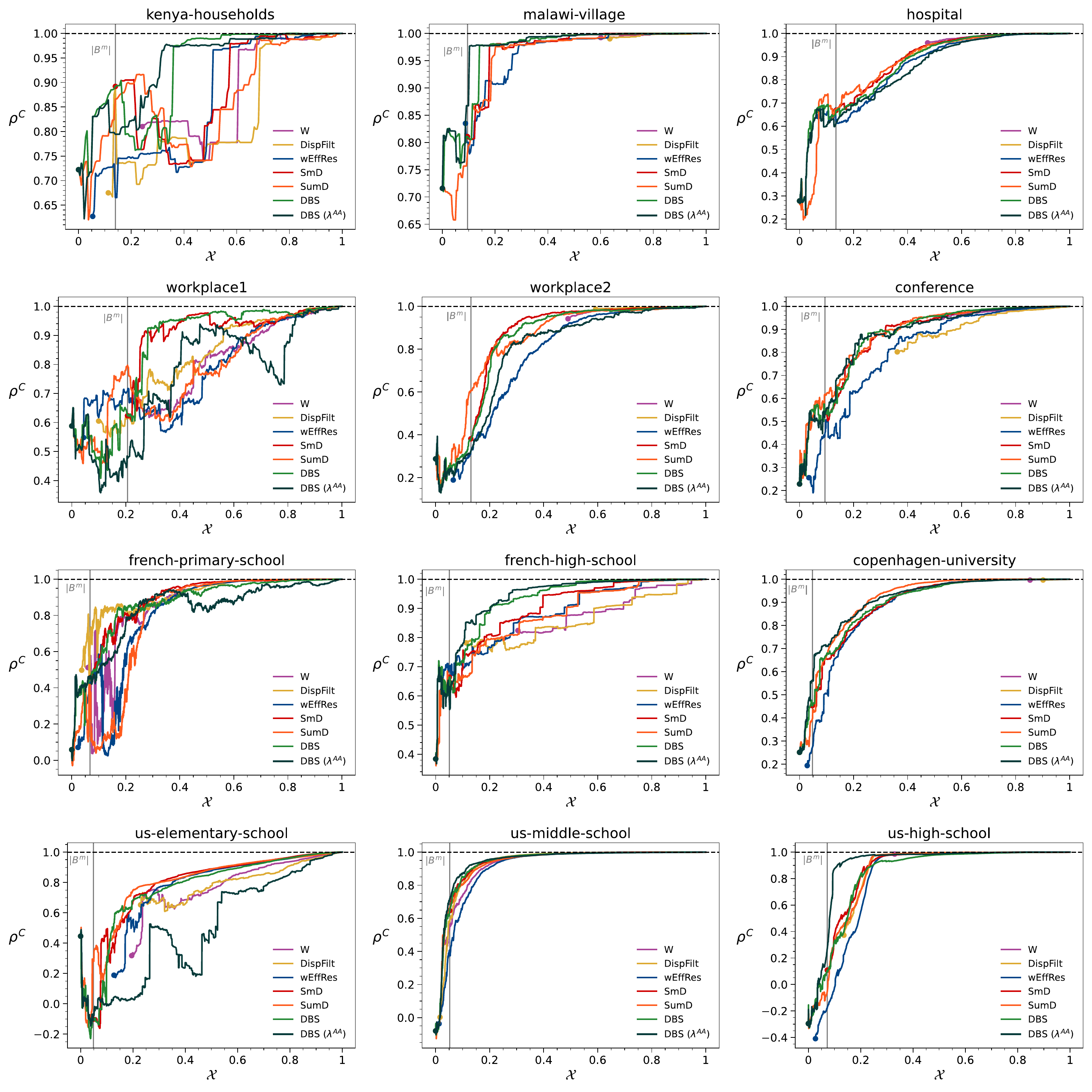}
\caption{Spearman's rank correlation between the eigenvector centrality of nodes in the sparsified network relative to the original network, $\rho^C$, for different levels of sparsification, $\mathcal{X}$. The curves do not contain error-bars for the measurements are deterministic. Each plot refers to one of the networks used in this study.}
\end{figure}

\newpage
\subsection{Preserving local dynamics: Infection Order}

\begin{figure}[h!]
\centering
\includegraphics[width=\linewidth]{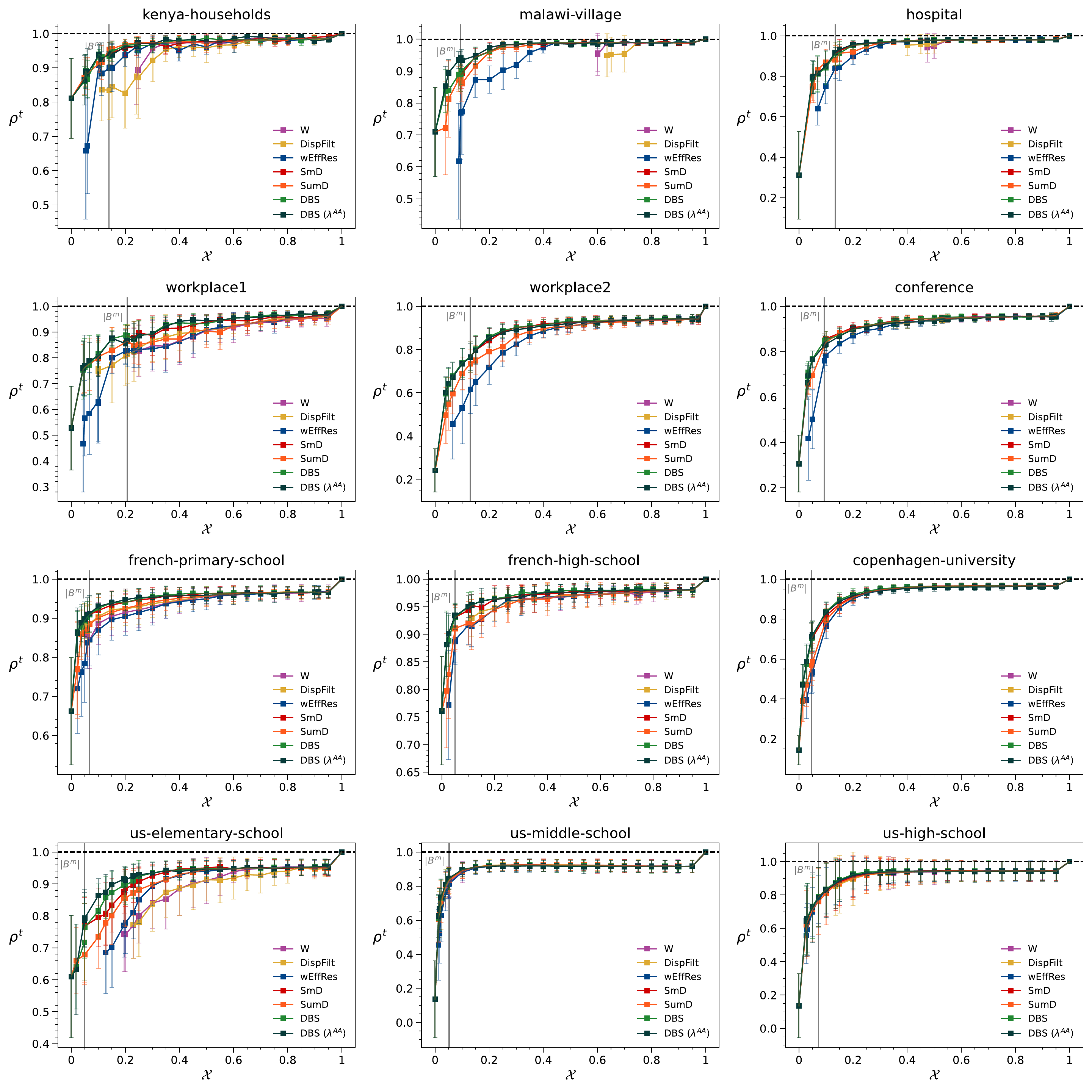}
\caption{Spearman's rank correlation between average time of infection, $\rho^t$, for different levels of sparsification, $\mathcal{X}$. Error-bars quantify the standard-deviation relative to choosing different seed nodes in the simulations. Each plot refers to one of the networks used in this study.}
\end{figure}

\newpage
\subsection{Preserving global dynamics: Time at which half of the nodes becomes infected}

\begin{figure}[h!]
\centering
\includegraphics[width=\linewidth]{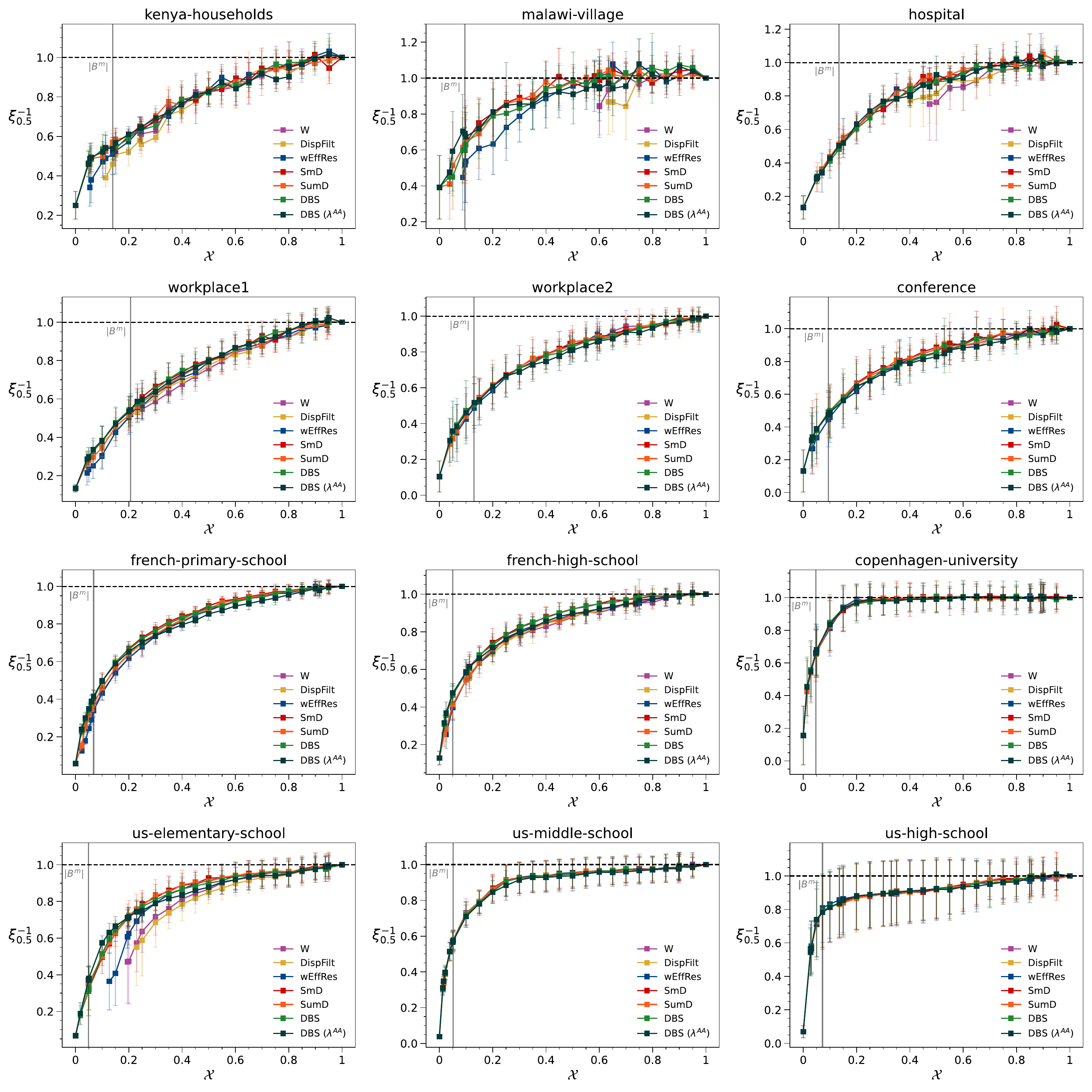}
\caption{Time at which half of the nodes becomes infected in the original network relative to the same time measured in the sparsified network, $\xi^{-1}_{0.5}$, for different levels of sparsification, $\mathcal{X}$. Error-bars quantify the standard-deviation relative to choosing different seed nodes in the simulations. Each plot refers to one of the networks used in this study.}
\end{figure}

\newpage
\subsection{Comparing with extra sparsification parameters}

\begin{figure}[h!]
\centering
\includegraphics[width=0.6\linewidth]{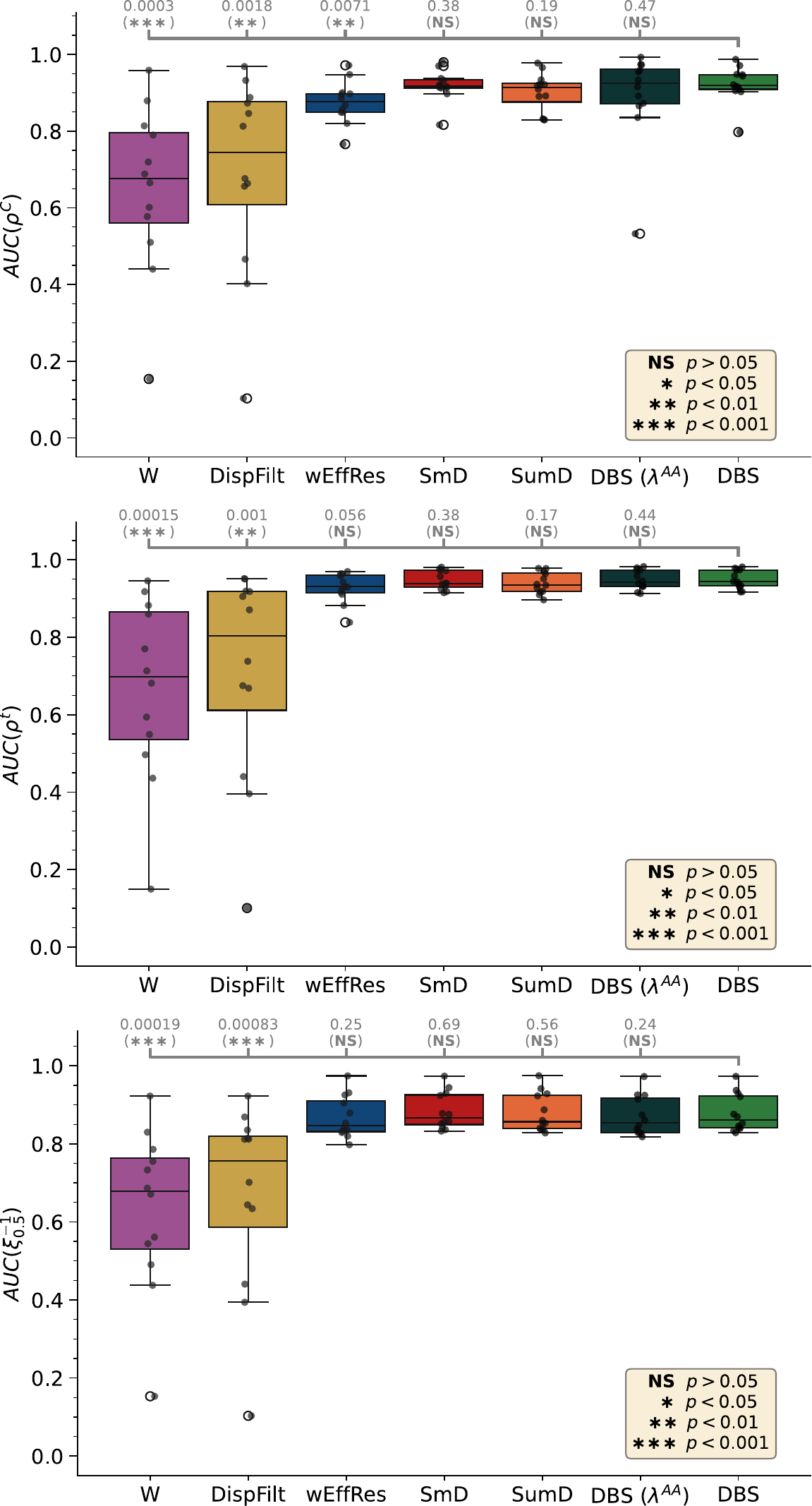}
\caption{AUC Boxplots considering the 12 Social Contact Networks. Comparison between different sparsfication schemes, \textbf{SumD}, \textbf{DBS($\lambda^{\text{AA}}$)} (using the Aczél-Alsina family of path-length measures) and \textbf{DBS} (using the Dombi family of path-length measures) as in the main text. \textit{(Top)} Boxplots respective to the AUC of the eigenvector centrality rank correlation curves, $\rho^C$. \textit{(Middle)} Boxplots respective to the AUC of the infection order correlation curves, $\rho^t$. \textit{(Bottom)} Boxplots respective to the AUC of the half-infection time proportion curves, $\xi^{-1}_{0.5}$.}
\end{figure}

\subsection{Dynamical performances are not affected by seed node}

In the main text, the spreading dynamics metrics were measured for $15\%$ of the nodes in each network as the seed nodes (where the infection started). 
This set comprises of $5\%$ of the most central, $5\%$ of the least central, and the other $5\%$ of the median central, all with respect to their eigenvector centrality in the original network.
Here we present the results for each of the three groups separately, showing that our results are robust to the choice of seed node.

\begin{figure}[h!]
\centering
\includegraphics[width=\linewidth]{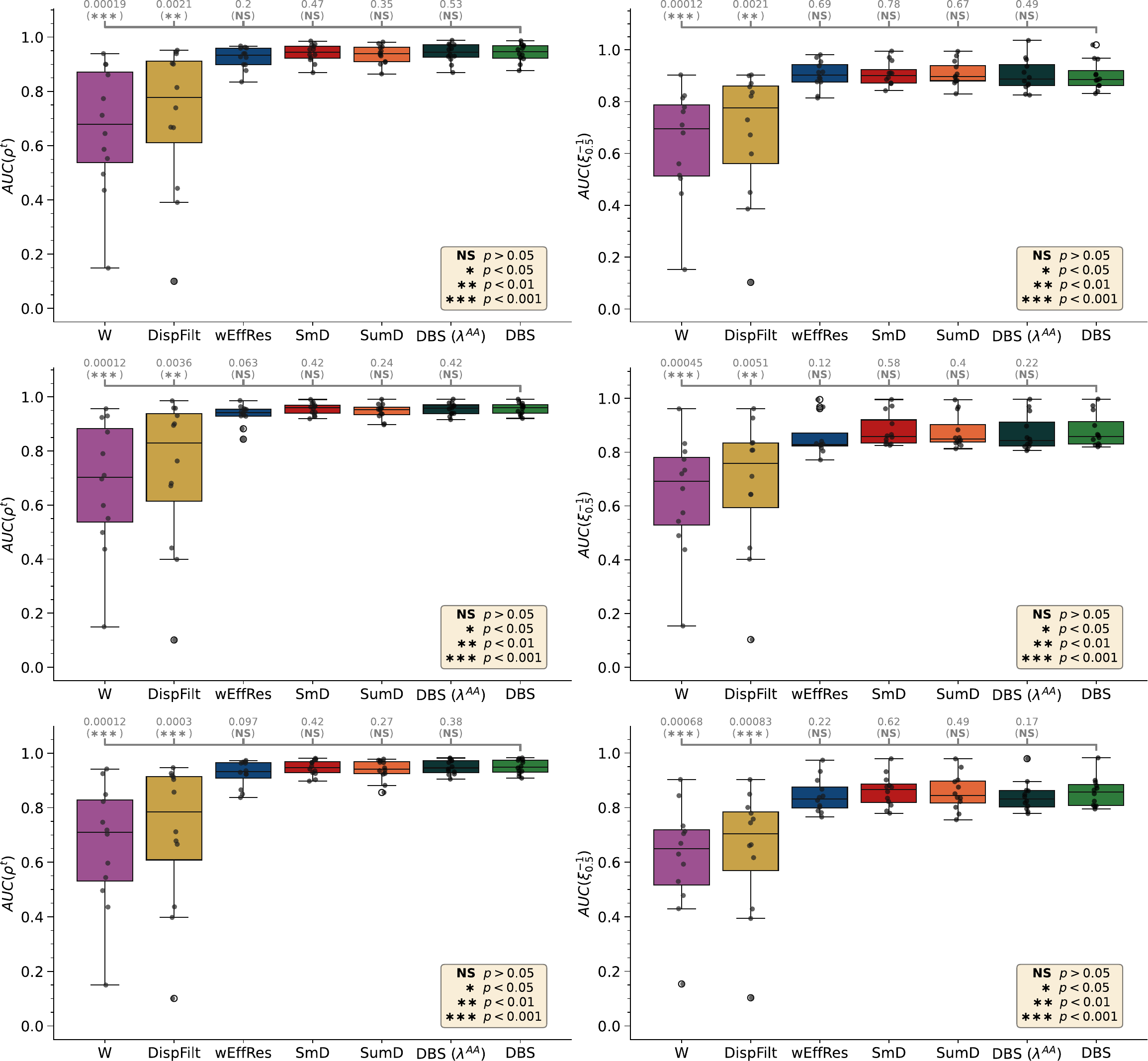}
\caption{Epidemic spreading related AUC boxplots, considering the 12 Social Contact Networks. Comparison between different sparsfication schemes, \textbf{SumD}, \textbf{DBS($\lambda^{\text{AA}}$)} (using the Aczél-Alsina family of path-length measures) and \textbf{DBS} (using the Dombi family of path-length measures) as in the main text. Data obtained using three subsets of seed nodes with different centralities, with respect to the eigenvector centralities of the original graph.
\textit{(Top Row)} \underline{5\% most central nodes}
\textit{(Middle Row)} \underline{5\% median central nodes}
\textit{(Bottom Row)} \underline{5\% least central nodes}
\textit{(Left Column)} Boxplots respective to the AUC of the infection order correlation curves, $\rho^t$. \textit{(Right Column)} Boxplots respective to the AUC of the half-infection time proportion curves, $\xi^{-1}_{0.5}$.}
\end{figure}

\subsection{Minimum subgraph sizes for specific performance requirements}

Another useful way of comparing the different sparsification schemes is by inspecting the minimum sizes of the outputted sparsified graphs ($\min \: \mathcal{X}$) that guarantee a given level of performance with respect to the original network.
In this case we investigated the necessary sizes of the sparsified subgraphs that would yield eigenvector centrality rank, $\rho^C$, infection order, $\rho^t$, and infection timing $\xi^{-1}_{0.5}$ above some threshold value (below we consider $0.9$, $0.7$, and $0.5$). 

\begin{figure}[h!]
\centering
\includegraphics[width=\linewidth]{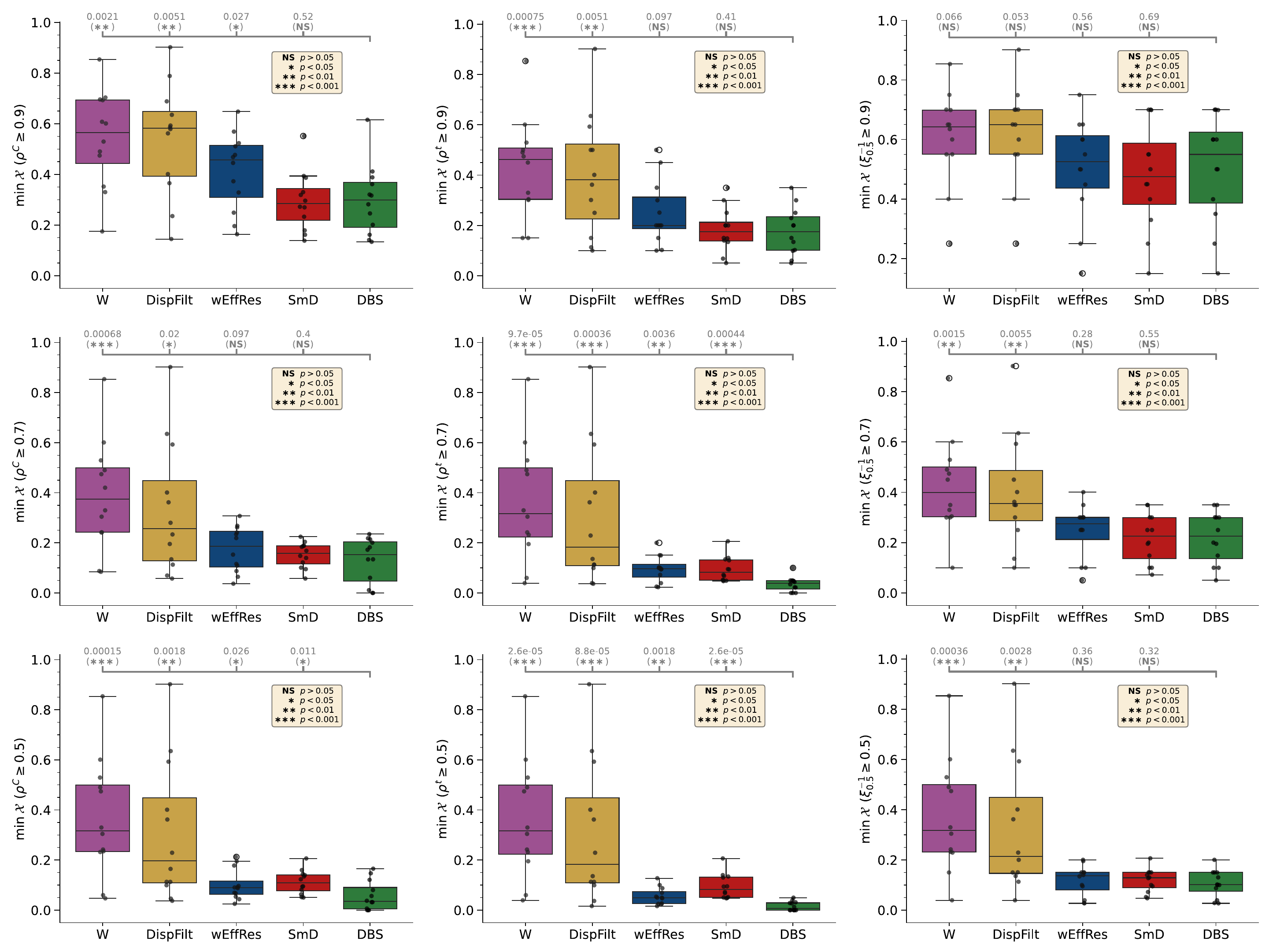}
\caption{Boxplots with the minimum sparsified subgraph sizes ($\min \: \mathcal{X}$) from each sparsification needed to attain values of $0.9$, $0.7$ and $0.5$ in the different performances measures, $\rho^C$, $\rho^t$ and $\xi^{-1}_{0.5}$ (from left to right). Results obtained using the 12 social contact networks.}
\label{fig:minimum_subgraph_size}
\end{figure}

\newpage
\subsection{Exhibit networks}
\label{subsec:exhibit}

Beyond the set of 12 social contact networks that we considered in the main text, we also inspected the performance of the former sparsification schemes using 69 networks gathered during an exhibit that took place at the Science Gallery of the Trinity College in Dublin, Ireland \cite{isella2011whats}.
The data collected in this event corresponds to close proximity events between people that randomly occurred during the exhibit which was open for 69 days \cite{sociopatterns2018}.
It interesting to note that the connectivity pattern of those networks are more random than more traditional social networks.
In those cases, the distance backbone synthesis (\textbf{DBS}) performs even better than other methods.
This set of networks has an average of $135.94 \pm 74.97$ nodes, $595.78 \pm 483.35$ edges, and a density of $0.079 \pm 0.068$.

\begin{figure}[h!]
\centering
\includegraphics[width=\linewidth]{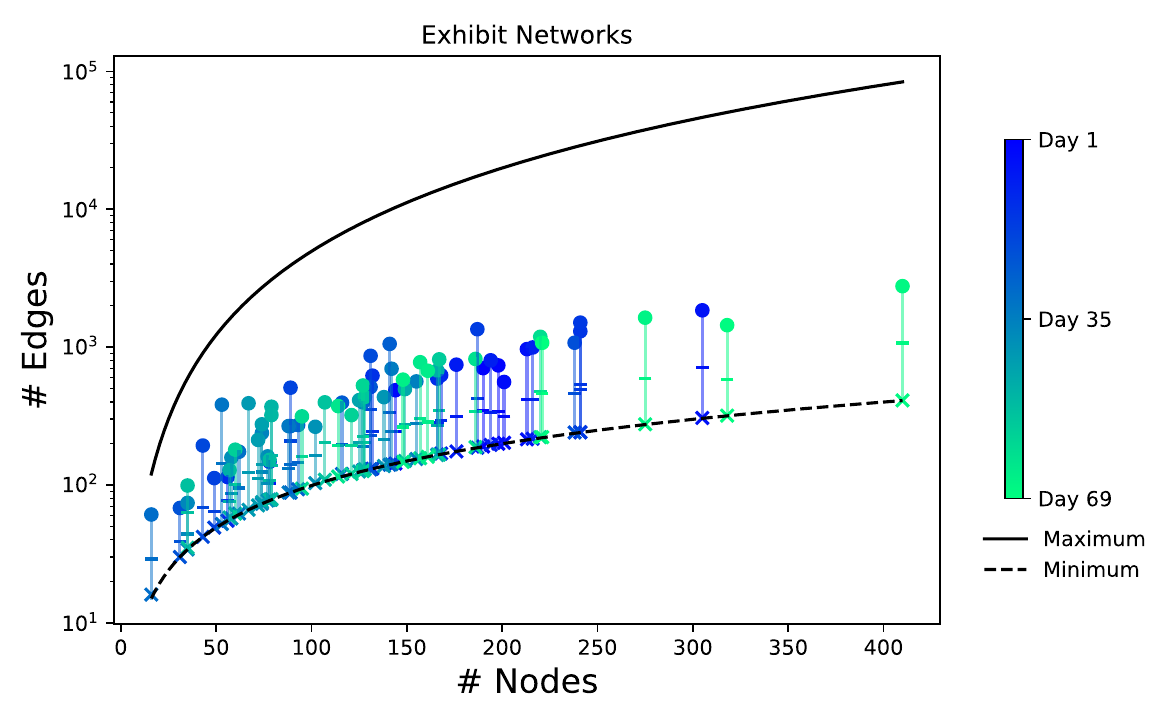}
\caption{Exhibit Networks Density. The black line corresponds to the maximum number of edges in a graph with a certain number of nodes and the dashed line corresponds to the smallest number of edges necessary to have a connected graph. The horizontal dashes in each network vertical colored line corresponds to the size of the metric backbone and the cross to the ultrametric backbone size.}
\end{figure}

In these networks, even the \textbf{DBS($\lambda^{\text{AA}}$)} and the \textbf{SumD} outperform every other method (with exception from semi-metric distortion) in every performance metric. Notably, \textbf{W} and \textbf{DispFilt} are clearly unsuited to sparsify networks with such a random structure, given the lower leaning distributions for both these methods. 

\newpage
\begin{figure}[h!]
\centering
\includegraphics[width=0.6\linewidth]{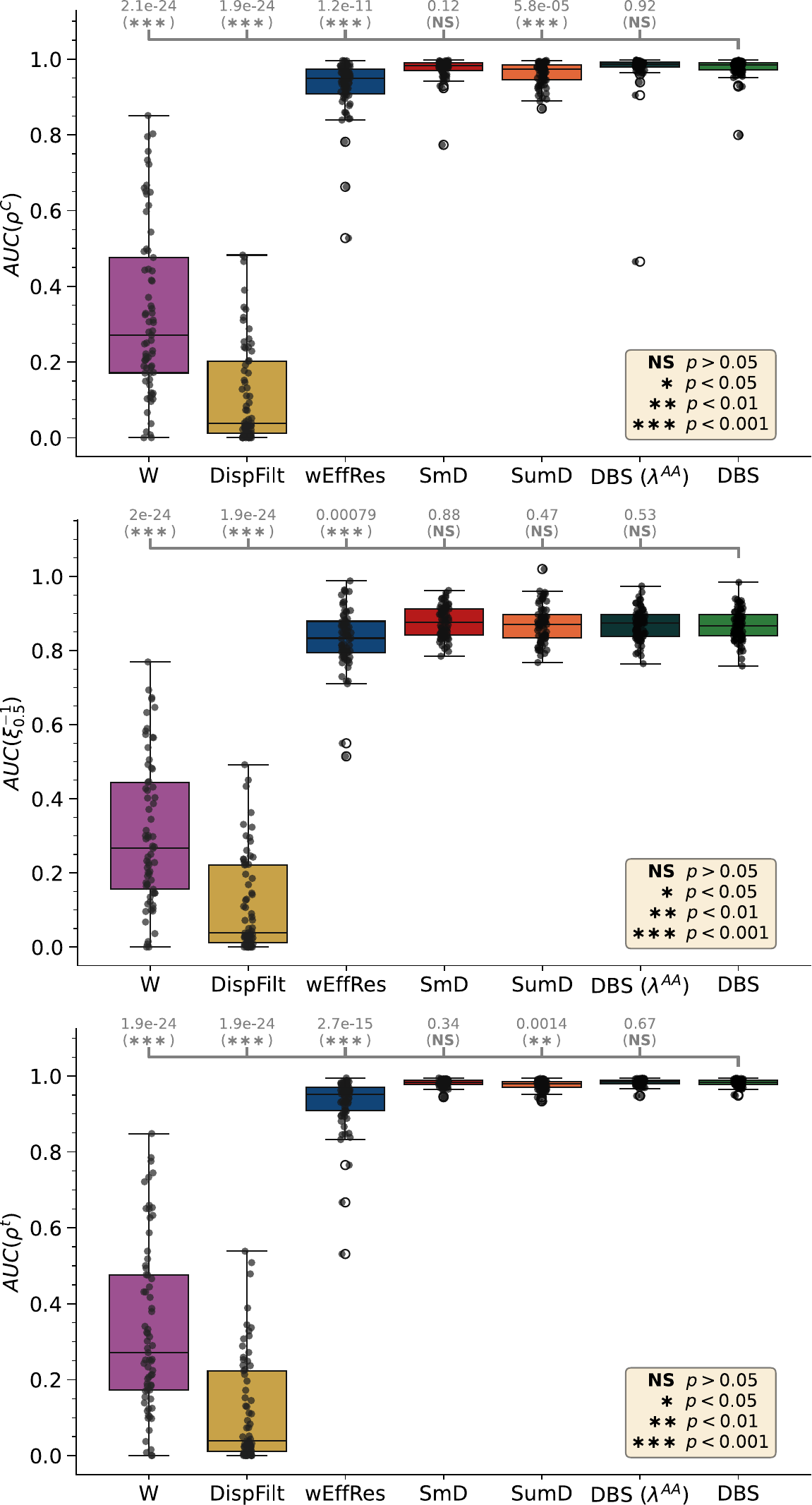}
\caption{AUC Boxplots considering the networks of the 69 days from the Exhibit Dataset. Comparison between different sparsfication schemes, \textbf{SumD}, \textbf{DBS($\lambda^{\text{AA}}$)} (using the Aczél-Alsina family of path-length measures) and \textbf{DBS} (using the Dombi family of path-length measures) as in the main text. \textit{(Top)} Boxplots respective to the AUC of the eigenvector centrality rank correlation curves, $\rho^C$. \textit{(Middle)} Boxplots respective to the AUC of the infection order correlation curves, $\rho^t$. \textit{(Bottom)} Boxplots respective to the AUC of the half-infection time proportion curves, $\xi^{-1}_{0.5}$.}
\end{figure}

\newpage
\subsubsection{Distance Backbone Synthesis performance is not affected by seed node}

\begin{figure}[h!]
\centering
\includegraphics[width=0.9\linewidth]{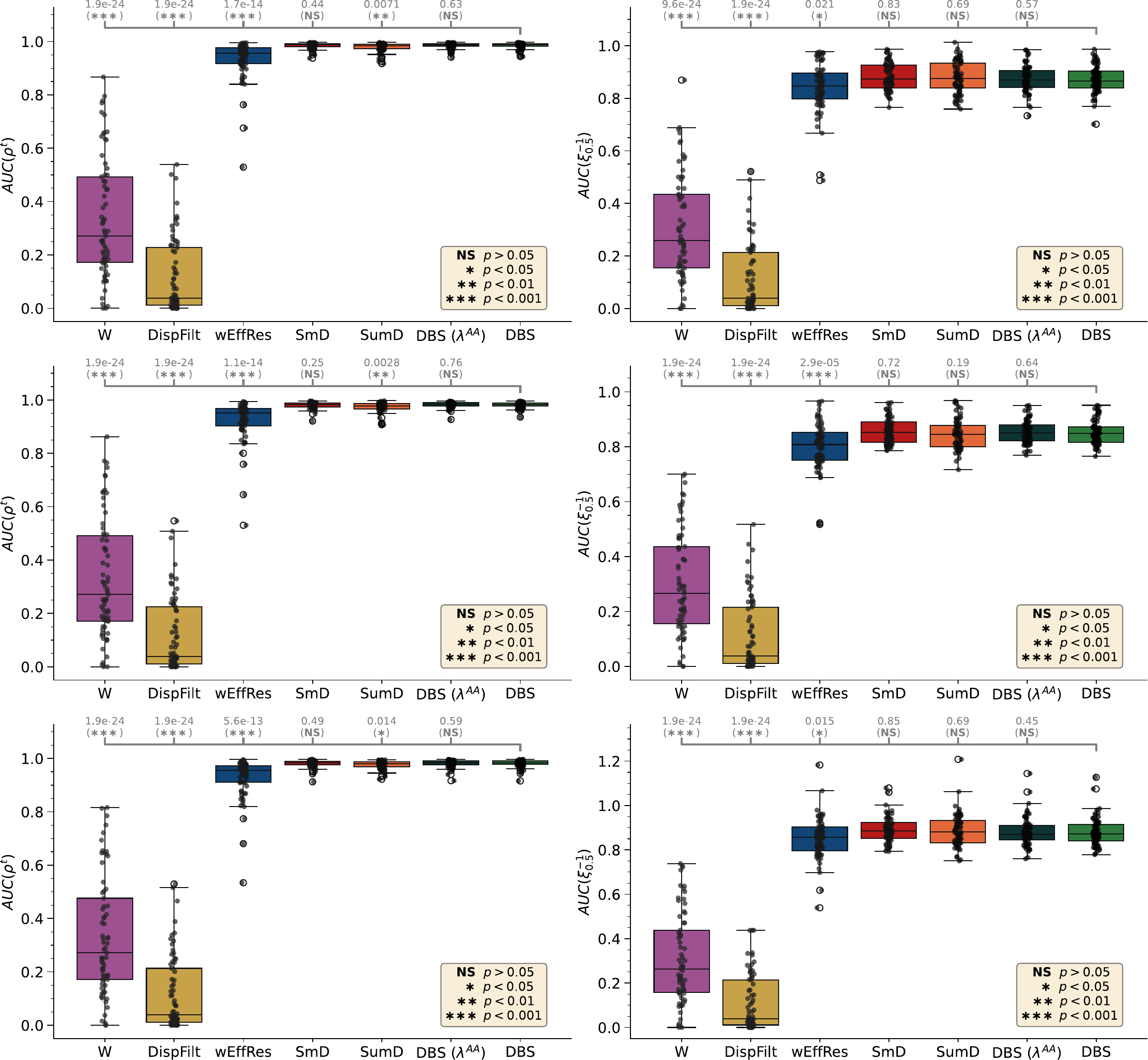}
\caption{Epidemic spreading related AUC boxplots, considering the 69 Exhibit Networks. Comparison between different sparsfication schemes, \textbf{SumD}, \textbf{DBS($\lambda^{\text{AA}}$)} (using the Aczél-Alsina family of path-length measures) and \textbf{DBS} (using the Dombi family of path-length measures) as in the main text. Data obtained using three subsets of seed nodes with different centralities, with respect to the eigenvector centralities of the original graph.
\textit{(Top Row)} \underline{5\% most central nodes}
\textit{(Middle Row)} \underline{5\% median central nodes}
\textit{(Bottom Row)} \underline{5\% least central nodes}
\textit{(Left Column)} Boxplots respective to the AUC of the infection order correlation curves, $\rho^t$. \textit{(Right Column)} Boxplots respective to the AUC of the half-infection time proportion curves, $\xi^{-1}_{0.5}$.}
\end{figure}

\newpage
\subsubsection{Minimum subgraph sizes for specific performance requirements}

\begin{figure}[h!]
\centering
\includegraphics[width=0.9\linewidth]{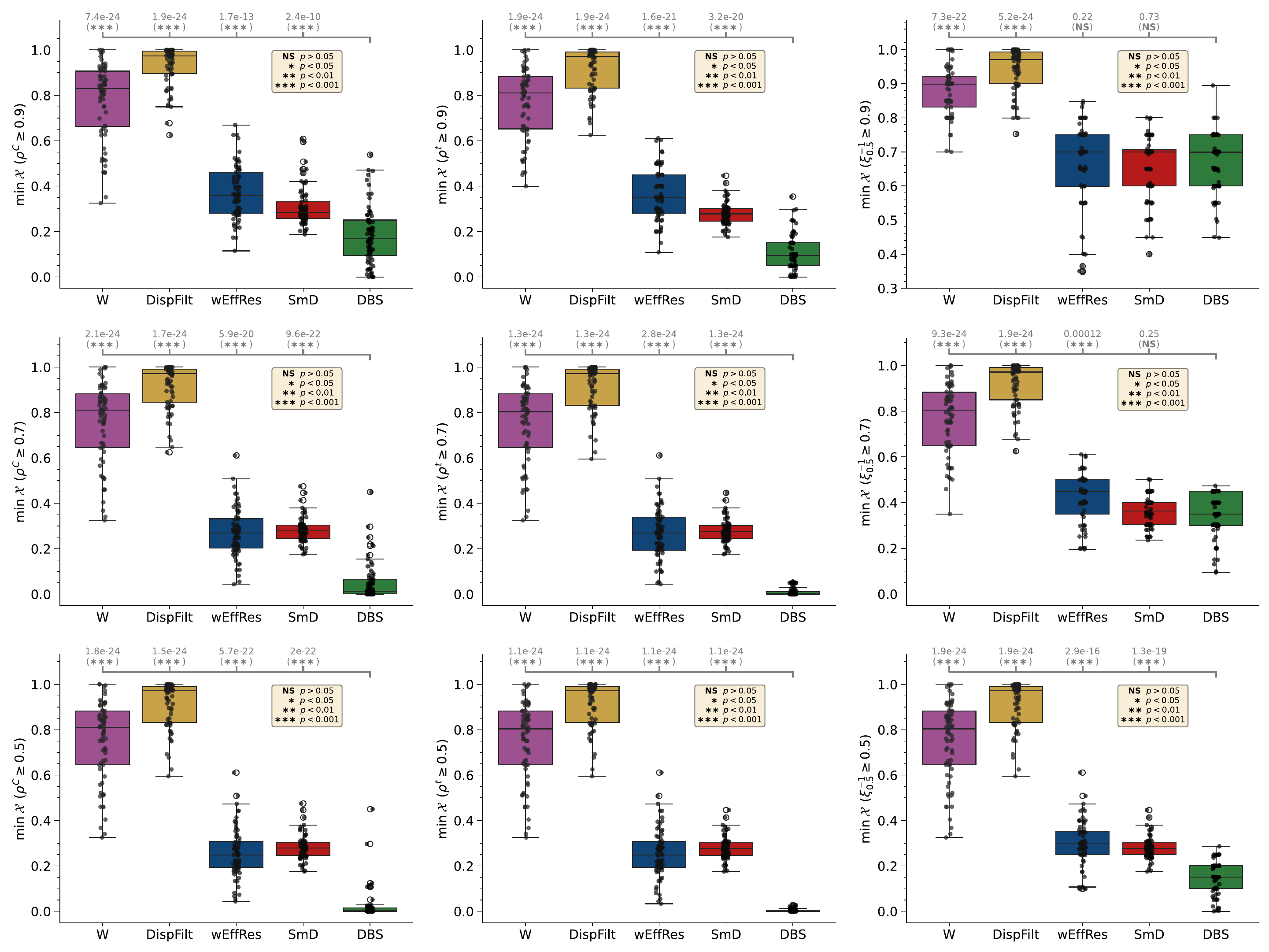}
\caption{Boxplots with the minimum sparsified subgraph sizes ($\min \: \mathcal{X}$) from each sparsification needed to attain values of $0.9$, $0.7$ and $0.5$ in the different performances measures, $\rho^C$, $\rho^t$ and $\xi^{-1}_{0.5}$ (from left to right). Results obtained using the 12 largest networks from the Exhibit dataset.}
\label{fig:minimum_subgraph_size_exhibit}
\end{figure}

\newpage
\subsection{Multiobjective optimization}

\begin{figure}[h!]
\centering
\includegraphics[width=\linewidth]{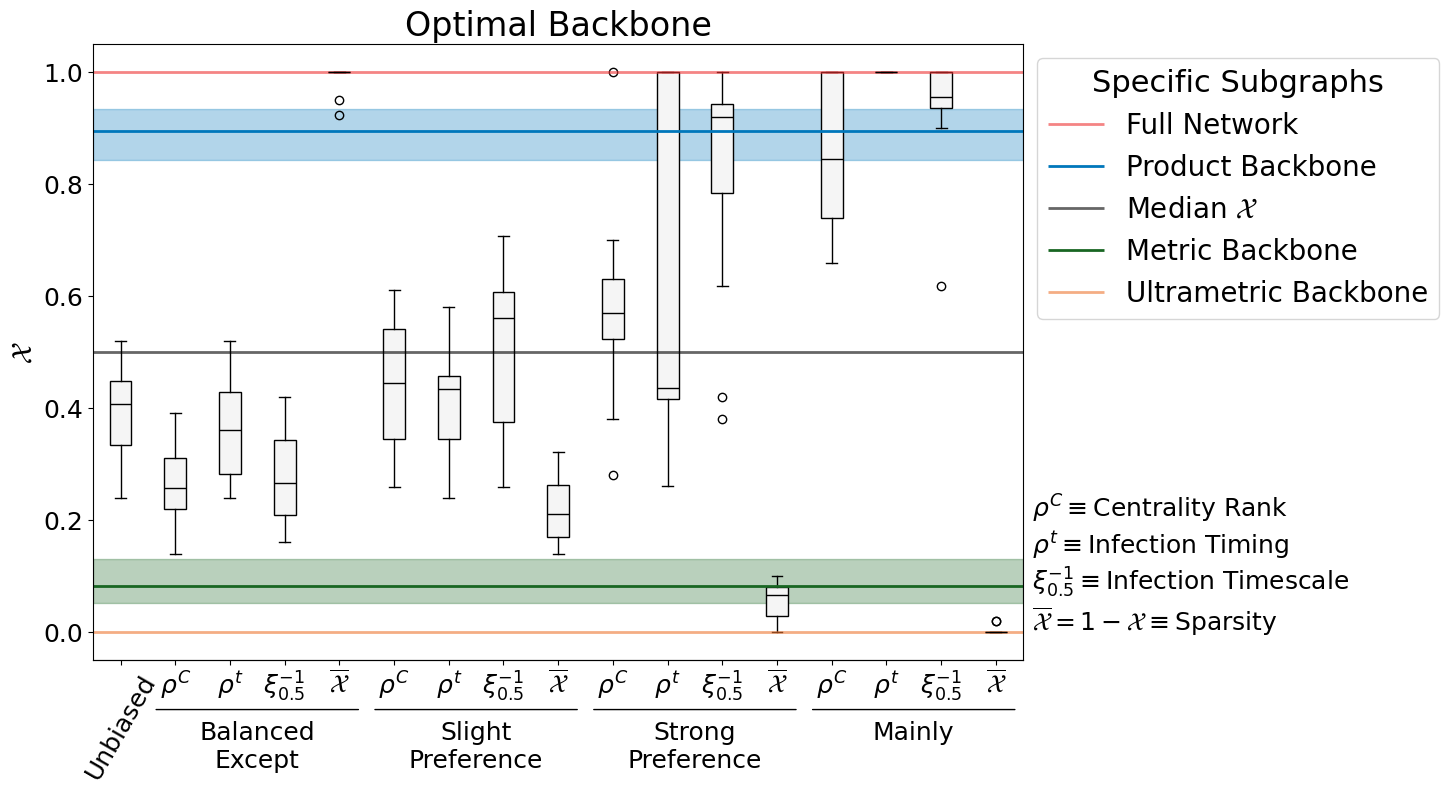}
\caption{\textbf{Optimal subgraph size that preserve desired network features.}
We measure the subgraph size $\chi$ for the optimal $\lambda^D$ found via a weighted sum of the eigenvector centrality ranking, $\rho^C$, infection timing ranking, $\rho^t$, infection timescale, $\xi^{-1}_{0.5}$, and sparsity, $1-\mathcal{X}$, over various scenarios described in Section~\ref{main:sec:optimization_methods} per network.
We plot the distribution of the optimal $\chi$ over the 12 social contact networks considered.
The particular cases of the metric backbone (light green) and the product backbone (light blue) are highlighted.
We also note the full network ($\chi=1$), the ultrametric backbone ($\chi=0$), and the median $\chi = 0.5$.
}
\label{fig:multi_objective_chi}
\end{figure}

\begin{figure}[h!]
\centering
\includegraphics[width=\linewidth]{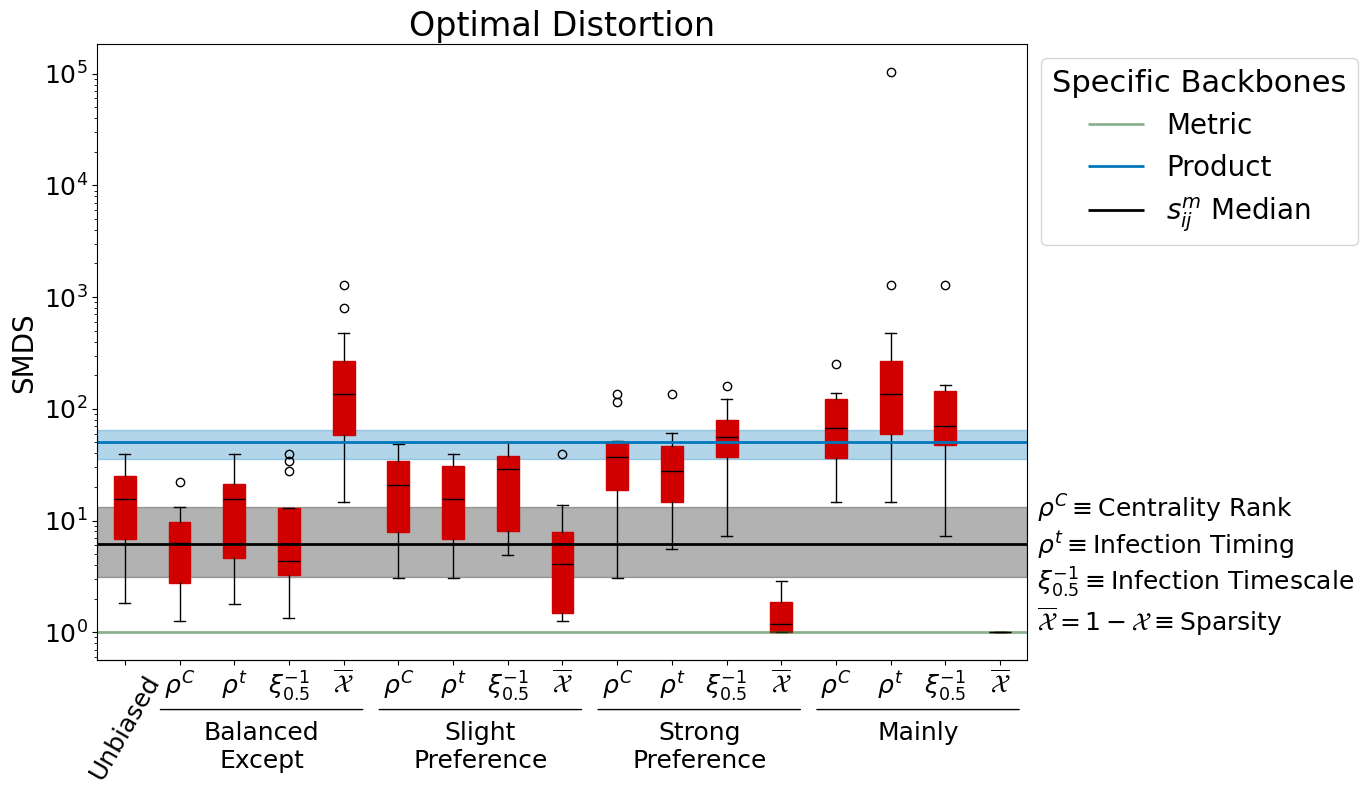}
\caption{\textbf{Optimal semi-metric distortion that preserve desired network features.}
We measure the optimal SMDS threshold found via a weighted sum of the eigenvector centrality ranking, $\rho^C$, infection timing ranking, $\rho^t$, infection timescale, $\xi^{-1}_{0.5}$, and sparsity, $1-\mathcal{X}$, over various scenarios described in Section~\ref{main:sec:optimization_methods} per network.
We plot the distribution of the optimal SMDS threshold over the 12 social contact networks considered.
The particular cases of the metric backbone (light green) and the product backbone (light blue) are highlighted.
The later is a small range of values for it can only be defined in a different parametrization, as discussed in Supplementary Section~\ref{sec:other_synthesis}.
We also shown the median and inter-quartile range, over the 12 networks, of the median $s_{ij}^m$ in black.
}
\label{fig:multi_objective_smds}
\end{figure}

\begin{figure}[h!]
\centering
\includegraphics[width=\linewidth]{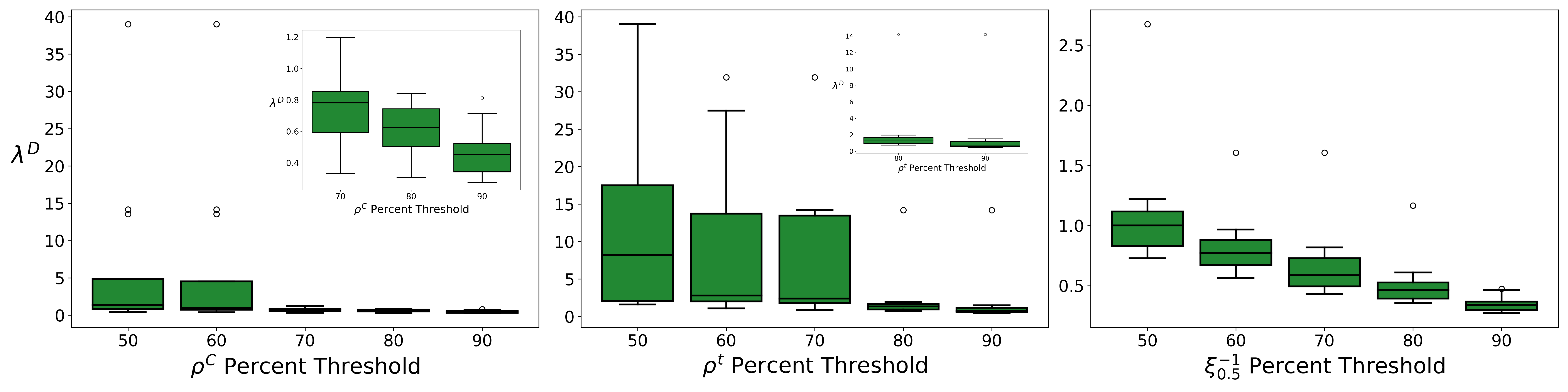}
\caption{\textbf{Distance Backbone Synthesis for specific performance requirements}
Value of the distance backbone synthesis parameter $\lambda^D$ for a specific percentage to recover eigenvector centrality ranking, $\rho^c$, and infection order, $\rho^t$, and timescale, $\xi^{-1}_{0.5}$.
We plot the distribution of $\lambda^D$ over the 12 social contact networks considered.
}
\label{fig:single_performance_lambda}
\end{figure}

\clearpage

\bibliographystyle{unsrt}  
\bibliography{references}